\documentclass[%
 	aip,
	jcp,
	superscriptaddress,
	reprint,
	amsfonts,
	amssymb,
	amsmath
]{revtex4-1}

\RequirePackage{amsthm}
\RequirePackage{calc}
\RequirePackage{color}

\RequirePackage{ifpdf}
\RequirePackage{graphicx}

\RequirePackage{ellipsis} 
\RequirePackage{ragged2e} 
\RequirePackage{microtype}

\input{QSE_JCP_latex_macros}
\input{QSE_JCP_defined_acronyms}

\newcommand{\myNewResult}[2][0.65\baselineskip]{%
	\vspace{#1}%
	\noindent{\scshape\mbox{#2\hspace{0.75em}}}%
}

\newcommand{\myCiteCanonical}{%
	\protect\cite{%
	Lee:2003ud,%
	Lee:1997uq,%
	Martin:02,%
	Arnold:199ye,%
	Abraham:1978uq,%
	Frenkel:1996ly,%
	Slichter:89,%
	Boehm:1989kx,%
	Nielsen:00,%
	Kloeden:1992kx,%
	Loan:96,%
	Greenbaum:1997nr%
}}

\newcommand{\mysubsubsection}[1]{\vspace{0.65\baselineskip}\noindent\textbf{#1}\vspace{0.45\baselineskip}\par}

\begin{document}

\title{%
Elements of naturality in dynamical simulation frameworks  \\
for Hamiltonian, thermostatic, and Lindbladian flows \\
on classical and quantum state-spaces 
\vspace{2ex}%
} 

\author{John A.\ Sidles}
\affiliation{Department of Orthopaedics and Sports Medicine, 
University of Washington}
\author{Joseph L.\ Garbini}
\affiliation{Department of Mechanical Engineering, University of Washington}
\author{Jonathan P.\ Jacky}
\affiliation{Department of Orthopaedics and Sports Medicine, 
University of Washington}
\author{Rico A.\ R.\ Picone} 
\affiliation{Department of Mechanical Engineering, University of Washington}
\author{Scott A.\ Harsila}
\affiliation{Department of Mechanical Engineering, University of Washington}

\collaboration{\rule[-2.25ex]{0pt}{4.75ex}UW Quantum System Engineering (QSE) Group}
\noaffiliation

\date{\today}

\begin{abstract}
The practical focus of this work is the dynamical simulation of polarization transport processes in quantum spin microscopy and spectroscopy.  The simulation framework is built-up  progressively, beginning with state-spaces (configuration manifolds) that are geometrically natural, introducing coordinates that are algebraically natural; and finally specifying dynamical potentials that are physically natural; in each respect explicit criteria are given for ``naturality.''  The resulting framework encompasses Hamiltonian flow (both classical and quantum), quantum Lindbladian processes, and classical thermostatic processes.  Lindbladian processes are shown to act generically to concentrate trajectories onto reduced-dimension state-spaces, such that the pullback of the symplectic, metric, and complex structure of Hilbert space---the {\Kahler} triple---is geometrically and informatically natural.  The~physical picture is that Lindbladian processes act to quench high-order correlations, such that the associated flow induces dimension-reduction.  A~concrete set of stochastic dynamical forms is provided, within both It\^{o} and Stratonovich formalisms, that is Lindblad-complete and suited to numerical computation.  These dynamical forms are shown to be the quantum counterpart of classical Langevin thermostats; they further encompass quantum measurement processes.   Constructive validation and verification criteria are given for metric and symplectic flows on classical, quantum, and hybrid state-spaces.  Two illustrative classical cases are developed: Riemannian geodesic trajectories on a torus, and rigid-body rotational dynamics of water molecules.  In particular, the projective naturality of a well-known quaternionic framework for simulating water molecule dynamics is proved as a special case of a general theorem that governs the construction of involutive subbundles upon canonically symplectic bundle manifolds. One quantum case is developed:  dynamic nuclear-spin polarization (\DNP) on tensor network state-spaces of varying rank. The resulting trajectory-oriented framework is well-suited to analyzing the practical performance and systems design challenges of classical, quantum, and hybrid systems like spin microscopes and spectrometers.\end{abstract}


\maketitle 

\onecolumngrid

\newpage

\twocolumngrid

\tableofcontents

\section{Introduction}
\label{sec: Introduction}

The work reported here has a practical focus: the end-to-end dynamical simulation of atomic-resolution quantum spin microscopy \cite{Sidles:2009yq}.  Here ``end-to-end'' means an integrated simulation of all the dynamical elements of a quantum spin microscope, from macroscopic elements like sample positioners, to mesoscopic elements like force microscope cantilevers, to fully quantum elements, like the individual spins in supramolecular structures.

Integrated simulations are necessary because present-day spin microscopes exhibit a level of complexity comparable to that of an earth-orbiting satellite.   Like satellites, spin microscopes are multisystem devices that are challenging to design, fabricate, operate, and repair.   Moreover, both satellites and spin microscopes seek to approach quantum and thermodynamic limits to performance as closely as feasible.    

To manage this complexity, it is routine engineering practice to simulate the performance of satellites (and comparably complex systems) end-to-end at every stage of  design, fabrication, and operation.  Our chief practical objective in this article is to provide a comparable quantum systems engineering capability to spin microscopy.

Among the most challenging aspects of systems engineering in spin microscopy is the quantum dynamics of the sample spins, which take place on a Hilbert state-space of enormous dimensionality. Even in the dynamically simpler context of magnetic resonance spectroscopy, existing quantum dynamical simulation methods typically scale poorly to large systems \cite{Bak:2000zp,Veshtort:2006fk}  or are ``plagued by uncontrolled approximations'' \cite{Greenbaum:2005uq}.    Spin microscopy introduces new dynamical elements---for example, ultra-strong magnetic field gradients---that further complicate the challenge of efficient quantum spin simulation.

Quantum systems engineers regard classical molecular dynamics simulations with envy for their applicability to large, inhomogeneous dynamical systems.  For example, it is routine practice to determine the diffusion constant $D$ of liquid water by integrating the trajectories of many thousands of water molecules.   The present article focuses upon creating a simulation framework having a parallel capacity for quantum spin dynamics, specifically, having the capacity to determine spin polarization transport coefficients by numerically integrating the dynamical trajectories of hundreds-to-thousands of quantum spins.

\begin{figure*}
{\centering%
\vspace{-0.1250\baselineskip}
\includegraphics[width=0.98\textwidth]{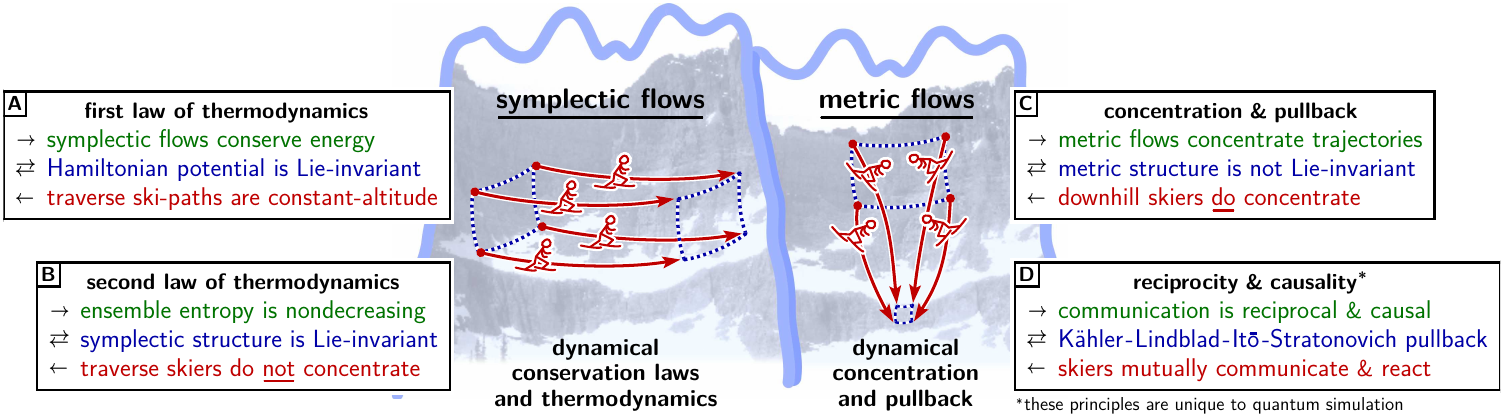}%
\caption[Geometric view of trajectory flow]{\label{fig: ski}%
Geometric perspective of trajectory flow in classical and quantum simulations. Dynamical flows are specified in two ways: as flow transverse to potential gradients (the~symplectic flow of \textbf{\sffamily A--B}) and as flow aligned with potential gradients (the~metric flow of \textbf{\sffamily C--D}); typically symplectic and metric flows are simultaneously present.  By~definition, a symplectic flow (\textbf{\sffamily A--B}) specifies the trajectory tangent vector from a potential gradient via a symplectic structure.  Symplectic flows  are associated with Hamiltonian processes; they support conservation laws and respect thermodynamic principles.   Metric flows (\textbf{\sffamily C--D}) specify the trajectory tangent vector from a potential gradient via a metric structure.  Metric flows are associated with thermostatic processes (both classical and quantum) and measurement processes (quantum only); metric flows concentrate trajectories and support pullback onto lower-dimension state-spaces. 
}}%
\end{figure*}

\subsection{Mathematical terminology}
\label{subsubsec: Mathematical terminology}

To assist readers, we italicize each mathematical term as it is introduced, and we follow Terence Tao's expository practice of providing some explanatory context at the first appearance of each mathematical term \cite{Tao:2008rr}.   For concision, though, we do \emph{not} define those mathematical terms that appear in the set of canonical texts \myCiteCanonical\  that we have listed in Sec.~\ref{subsec: Dovetailed elements of naturality}, and we refrain almost entirely from introducing new mathematical terms.\footnote{The sole new mathematical terms introduced in this article are the definitions of \emph{$F$-compatible} and \emph{$F$-natural} simulation metrics that appear in Theorems 1--3 of Tables~\ref{table: metric flow}--\ref{table: symplectic flow}, and the mnemonic \protect\KLISP, which is an acronym for ``\protect\Kahlerian-Lindbladian-\Ito-Stratonovich pullback,'' that appears in Table~\ref{table: summary}.}  For us, the terms \emph{Hamiltonian} and \emph{Lindbladian} serve as both nouns and adjectives, and in particular the Lindbladian is a set of \emph{stochastic potentials} (see Table~\ref{table: recipes}), originating in quantum information theory, that encompasses and generalizes the notion of a Hamiltonian.  Our framework reduces simulation to the computation of \emph{trajectories} as \emph{integral curves} of Lindbladian potentials.

\subsection{Overview of dimension reduction}
\label{subsubsec: Overview of dimension reduction}

As is common in the literature, in this article we take \emph{reduction} to be a generic name for a constellation of mathematical tools that, given a dynamical system, ``restricts the study of its dynamics to a system of smaller dimension.'' \cite{Ortega:2005uq}  Particularly in the numerical simulation of quantum systems---for which the native \emph{Hilbert space} has exponentially many dimensions---dimension-reduction is a practical necessity.

Our approach to reduction focuses on thermodynamical processes; thus our framework is best suited to systems that are (or have been) in contact with a thermal reservoir.  In classical dynamics thermalizing processes are commonly called \emph{thermostats}; in quantum dynamics they are commonly called \emph{Lindbladian processes}.   

In this article we regard Lindbladian processes as being fundamental, such that classical thermostatic processes emerge in a (well posed) quantum-to-classical limit. Hamiltonian potentials will govern symplectic dynamical flows, as usual in both classical and quantum physics, but even these Hamiltonian potentials will be subsumed into Lindbladian processes.

Turning to a geometric point-of-view, state-space trajectories can be regarded as arising alternatively from \emph{metric structure} or from \emph{symplectic structure}; we will find that thermodynamical flows are metric, while Hamiltonian flows are symplectic.  

We therefore require of our simulations---both classical and quantum---that the state-spaces be endowed with {both} metric and symplectic structure. For thermodynamic processes in particular (and also for control processes) the dynamical potentials of our simulations will be \emph{stochastic}, and we will see that the stochastic nature of our potentials acts both to thermalize the dynamics and to enforce the standard quantum limits to measurement processes and backaction.

We will find that Lindbladian reduction is well-suited to simulating large-scale quantum dynamical systems that lack symmetry (or in which the symmetry is present but unrecognized), but which are in contact with a thermal reservoir; quantum spin biomicroscopy is a paradigmatic example of this class of systems.

\subsection{Lindbladian versus Hamiltonian reduction}
\label{subsec: Lindbladian versus Hamiltonian reduction}

Symmetries of Hamiltonian systems have long been known to enable dimensional reduction, and this has led to a constellation of closely related dimension reduction methods that are variously called \emph{Hamiltonian}, \emph{symplectic}, \emph{symmetric}, \emph{geometric}, or \emph{Lie reduction} \cite{Ortega:2005uq,Ortega:2004fk,Marsden:2007fk,Ortega:2004kx}.  These various names reflect, not varying methods, but varying emphasis: ``Hamiltonian'' as a modifier emphasizes the role of dynamical potentials, ``symplectic'' emphasizes that the dynamical flow is a symplectomorphism, ``symmetric'' emphasizes the central role of symmetry in the dimension reduction, ``geometric'' emphasizes considerations relating to differential geometry, and ``Lie'' emphasizes the central role that Lie groups play in these methods.

Our simulation framework comprises an alternative approach to dimension reduction that can variously be called \emph{Lindbladian}, \emph{metric}, \emph{thermostatic}, or \emph{synoptic}.  Here ``Lindbladian'' emphasizes the key role of quantum stochastic processes, ``metric'' emphasizes the role of Riemannian metric flow, ``thermostatic'' emphasizes reduction arising from dynamical interaction with a thermal bath, and ``synoptic'' emphasizes that Lindbladian processes associated with thermalization can always be described as equivalent continuous measurement processes that synoptically observe the dynamics \cite{Sidles:2009cl}.  

Our preferred phrase will be ``Lindbladian reduction'' in recognition of the key physical idea is that dynamical flows associated with noise, thermalization, or measurement (Lindbladian processes accommodate all three) act generically to compress trajectories---both classical and quantum---onto state-spaces of lower dimension.  

\subsection{New results in this article}
\label{subsec: New results in this article}

We now list those of our results that are new (to the best of our knowledge), together with the sections in which they are derived.

\myNewResult{Lindbladian reduction}%
The expressions given in Appendix \ref{sec: Lindbladian recipes} and in Table~\ref{table: recipes} express general Lindbladian processes in terms of stochastic flows that are specified in terms of potentials known as \emph{Berezin symbols}.

\myNewResult{Compatible structures}%
Theorems 1--3 of Sec.~\ref{sec: The summit: valid design criteria} and Tables~\ref{table: metric flow}--\ref{table: symplectic flow} link metric structure compatibly with symplectic structure.

\myNewResult{Bloch potentials}%
The \emph{Bloch equations} provide a much-used paradigmatic description of spin thermalization; in Appendix \ref{sec: Lindbladian recipes} we derive a complete set of Bloch-equivalent Lindbladian potentials.

\myNewResult{Factored metric}%
The metric structure specified in Table \ref{table: summary}(J--L) is given in a factored representation that is enabling for large-scale simulation codes.  

\myNewResult{Validation and verification criteria}%
Large-scale simulation codes are themselves engineered objects; the validation and verification criteria of Table~\ref{table: metric flow}(E) and Table~\ref{table: symplectic flow}(F) are tools for creating these codes.

\subsection{Dovetailed elements of naturality}%
\label{subsec: Dovetailed elements of naturality}

We have organized our framework around dovetailed notions of mathematical naturality.  For convenience we have identifoed a set of twelve canonical textbooks \myCiteCanonical\ that, in aggregate, convey all of the elements of naturality that we require.   We review them in a recommended reading order, which we also adopt as a descending order of precedence for mathematical notation.  Thus the ideas of mathematical naturality that are most important to our simulation framework are systematically given the highest notational precedence.

\myNewResult{Geometric naturality}%
We take geometric naturality to have the highest mathematical precedence, so that the elements of our simulation framework are specified in geometric terms whenever possible. The key elements of geometric naturality are taken to be the \emph{pullback} of dynamical potentials to the simulation state-space, the specification of \emph{compatible} metric and symplectic structure, and the \emph{pushforward} of trajectories to the native state-space.  We adopt the notation and terminology of Lee's \emph{Smooth Manifolds} \cite{Lee:2003ud} and \emph{Riemannian Manifolds} \cite{Lee:1997uq}, with minimal extensions to complex manifolds as described in Martin \cite{Martin:02}.  With one necessary exception (see the heading ``algebraic naturality'' below) the elements of our simulation framework are compatible with geometrically natural constructions; this allows the development of simulation software to be largely automated.  

\begin{figure}
\centering%
\vspace{-0.1250\baselineskip}
\includegraphics[width=0.98\columnwidth]{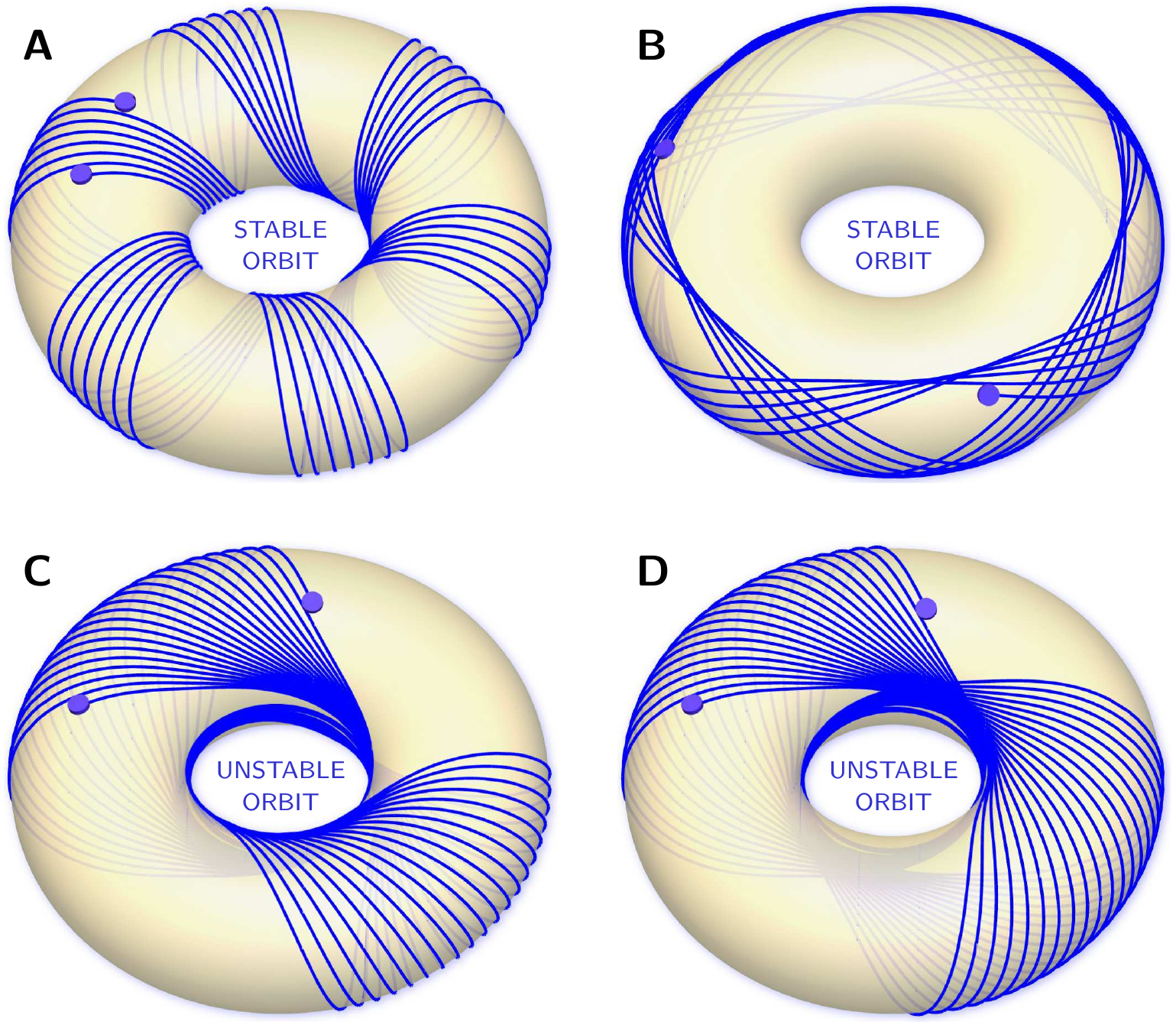}%
\caption[Geodesic trajectories on a torus]{\label{fig: torus}%
Geodesic trajectories on a torus.  The trajectories are numerically integrated upon an algebraic state-space that is constructed via Theorem~1 (Compatible Metric Structure) and Theorem~2 (Compatible Symplectic Structure); see Tables~\ref{table: metric flow}--\ref{table: symplectic flow} and Sec.~\ref{subsec: introducing coordinates}. (\textbf{\sffamily A--B})\hspace{0.4em}convergent geodesic trajectories mainly traverse regions of positive Riemannian curvature, while divergent geodesic trajectories (\textbf{\sffamily C--D})\hspace{0.4em} mainly traverse regions of negative curvature (see \onlinecite[Lee][Ch.~10]{Lee:1997uq} for details).  The starting orientation of trajectory (\textbf{\sffamily C}) differs from those of (\textbf{\sffamily D}) by $\sim1.0{\times}10^{-10}$ radians; both trajectories circle the central hole six times before returning near to their starting point; the (\textbf{\sffamily C}) trajectories thread the central hole and the(\textbf{\sffamily D}) trajectories don't.   Algebraic manifolds are well-suited to the high-accuracy, numerically efficient trajectory integration that is required to discriminate these fine differences. }%
\end{figure}

\myNewResult{Dynamical naturality}%
The key elements of dynamical naturality we take to be a symplectic flow governed by Hamiltonian potentials, and metric flow that describes thermalizing dynamics and measurement processes.  We adopt the notation and terminology of Arnold \cite{Arnold:199ye} and of Abraham and Marsden \cite{Abraham:1978uq}, with extensions for classical thermostatic dynamics from Frenkel and Smit \cite{Frenkel:1996ly}.  Magnetic resonance dynamics, and in particular Bloch relaxation, are specified in the notation of Slichter \cite{Slichter:89}.  Dynamical naturality is integrated with geometric naturality by the geometrically natural pullback of dynamical potentials, metric structure, and symplectic structure.

\myNewResult{Projective naturality}%
Combining the preceding notions of geometric and dynamical naturality leads us to a notion of \emph{projective naturality} that is concretely realized as the validation and verification criteria of Table~\ref{table: metric flow}(E) and Table~\ref{table: symplectic flow}(F), which we construct by methods that are geometrically natural.  The key role in systems engineering of validation and verification criteria are described in a classic article of Boehm \cite{Boehm:1984fk,Boehm:1989kx}; projective naturality amounts to the expression of Boehm's criteria in a form that is geometrically and dynamically natural.

\myNewResult{Informatic naturality}%
In our framework all processes of thermalization, measurement, and control are realized as stochastic Lindbladian flows, which we develop by merging the quantum informatic approach of Nielsen and Chuang (\onlinecite[chs.~2 and 8]{Nielsen:00})  with the stochastic approach of Kloeden and Platen \cite{Kloeden:1992kx}.  Nielsen and Chuang specify Lindbladian dynamics in an algebraic form that our simulation framework translatse into one-forms specified by Berezin symbols (see Appendix \ref{sec: Lindbladian recipes} for recipes; see (\onlinecite[][sec.~3 and figs.~8--9]{Sidles:2009cl}) for physical systems that exhibit Lindbladian dynamics). The resulting general Lindbladian  is natural in geometric, dynamical, and projective senses described above; this too helps in automating the development of simulation software.

\myNewResult{Algebraic naturality}
For reasons of simulation efficience, we design the symplectic and metric structure of our simulations to have an algebraic structure that we associate to a choice of algebraic coordinates (see Sec.~\ref{subsec: constructing a natural simulation}).  The specification of algebraic coordinates is the sole aspect of our framework that is not natural the geometric, dynamical, projective, and informatic senses described above.  Once coordinates have been specified, the ensuing trajectory integration is carried through by standard numerical computation methods that are adapted largely from Golub and Van Loan \cite{Loan:96} and from Greenbaum \cite{Greenbaum:1997nr}.

\myNewResult{Further aspects of naturality}
\label{subsubsec: Other aspects of naturality}
Our framework focuses on the simulation of noisy, asymmetric systems; we therefore have little occasion to refer to the literature on symmetric reduction methods \cite{Ortega:2005uq,Ortega:2004fk,Marsden:2007fk,Ortega:2004kx}, despite the undoubted utility of symmetry reduction in particular cases.   

Similarly, our quantum simulation state-spaces in general are not Hilbert spaces, but rather are lower-dimension nonlinear tensor network manifolds.  Thus we will have little occasion to apply such familiar tools of quantum mechanics as density matrices or projective measurements, because these are not defined on tensor network manifolds.   

Our own recent analysis \cite{Sidles:2009cl} provides derivations of many of the standard results of quantum mechanics---including uncertainty principles, standard quantum limits to noise and measurement accuracy, and projective measurements---upon this non-Hilbert quantum state-space.  This article is recommended to readers who wish to work through the details of how results typically derived on linear Hilbert spaces can be alternatively derived via a Lindbladian formalism on tensor network manifolds. 

\begin{figure}
\centering%
\vspace{-0.1250\baselineskip}
\includegraphics[width=0.98\columnwidth]{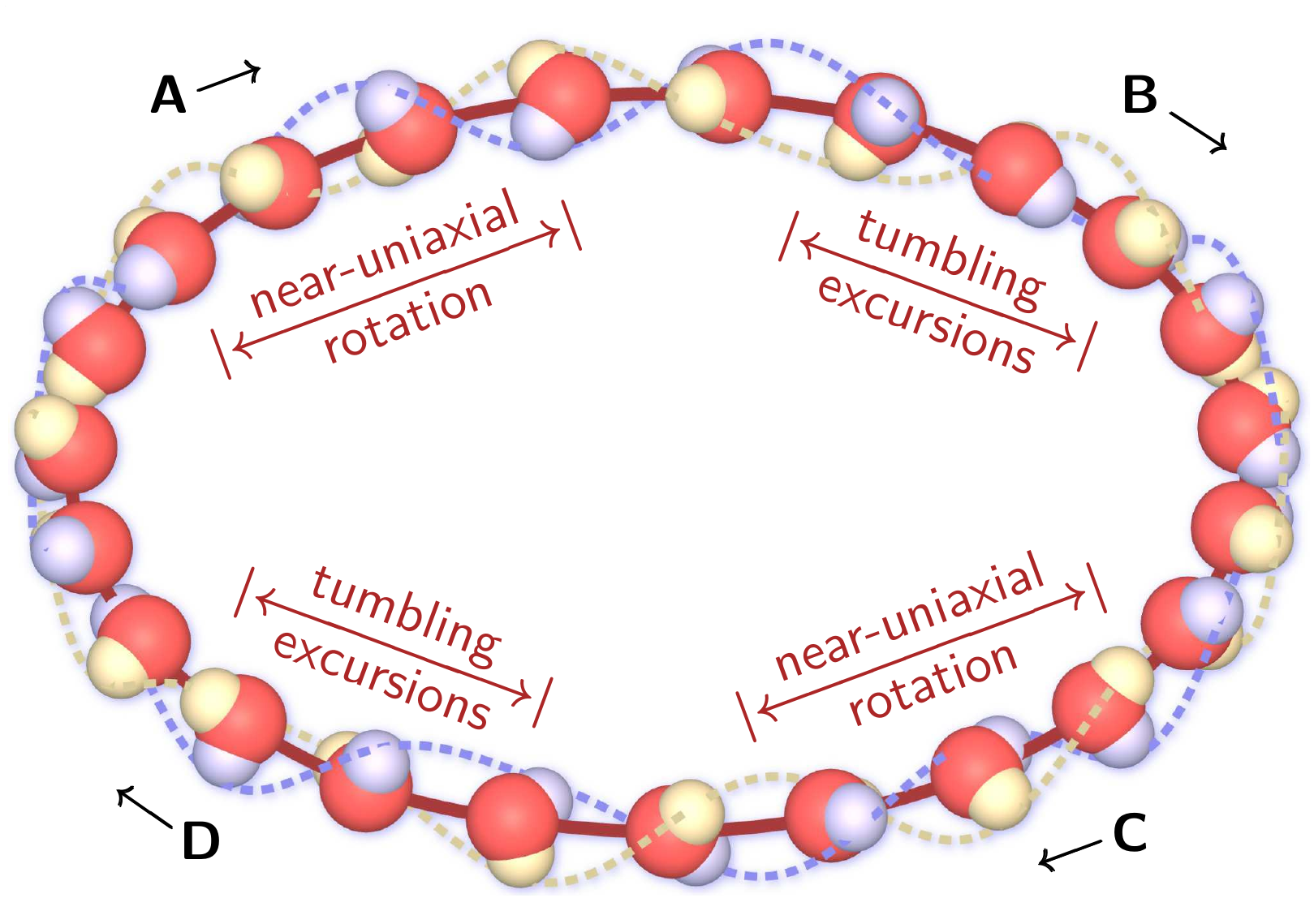}%
\caption[Rotation of a water molecule]{\label{fig: rigid}%
Rigid body rotation of a water molecule, accompanied by center-of-mass motion in a harmonic potential.  As with the toric trajectories of Fig.~\ref{fig: torus}, this motion is formally a geodesic trajectory upon an algebraic state-space that is constructed via Theorems~1--3 of Tables~\ref{table: metric flow}--\ref{table: symplectic flow}.  The rotational sector of the state-space so constructed is essentially identical to the quaternionic state-space of {Miller \emph{et al{.}} \cite{Miller-III:2002fk}}; see Sec.~\ref{subsec: rigid body rotation} for details.  The molecule's angular momentum is initialized to be nearly aligned with the \HtwoO\ molecule's middle moment of rotational inertia; in consequence the rotational motion consists of intervals of stable precession alternating with intervals of rotational tumbling.  Astronaut Michael Foale has given a diverting account of the practical consequences of this long-known tumbling instability \cite{Foale:1999fk}.%
}%
\end{figure}

\myNewResult{Educational objectives}  It plausible that during the last two centuries, more people have learned the basic principles of differential geometry from Nathaniel Bowditch's nautical textbook \emph{The American Practical Navigator} (commonly known as ``\emph{Bowditch}'') than from any other single book or article.  It is less widely appreciated, however, that editions of Bowditch as early as 1807 are startlingly rigorous in their mathematical description of what became known as Riemannian geometry (\onlinecite[][p.~91]{Bowditch:1807fk}), decades in advance of Gauss' pioneering work. 

What accounts for the enduring success of \emph{Bowditch} over the past two centuries?  There is a story \cite{Bowditch:2002fk} that\begin{quote}
\ldots\,Bowditch vowed while writing (the first American) edition to ``put down in the book nothing I can't teach the crew,'' and it is said that every member of his crew including the cook could take a lunar observation and plot the ship's position.
\end{quote}Thus  \emph{Bowditch} provides a vivid historical example of Mac Lane's point that ``Mechanics developed by the treatment of many specific problems'' (\onlinecite[][p.~295]{Mac-Lane:1986qe}) and the observation of Abraham and Marsden that ``In mechanics almost every branch of mathematics gets used'' (\onlinecite[p.~15]{Abraham:1978uq}).  The three worked examples in this article were crafted, therefore, with conscious attention to the principles of Bowditch, Mac Lane, Abraham, and Marsden.

\myNewResult{Scientific objectives}
As we gain a broader appreciation that the dynamical elements of quantum mechanics survive pullback onto nonlinear state-spaces, the natural question arises ``Is the state-space of quantum mechanics a linear Hilbert space, or does it have some more general---even dynamical---geometry?''  Formally speaking, in this article we view the state-space of nature as a Hilbert space, to which our simulation framework provides a useful approximation.  There is a substantial body of literature, however, that takes the opposite point-of-view, which we review in posing  Challenge~\ref{challenge: oracles} of Sec.~\ref{subsec: Demonstrate quantum oracles}.

\section{The ascent: geometric dynamics}
\label{sec: The ascent: geometric dynamics}

We start with the idea that Hamiltonian simulation frameworks---whether classical or quantum---can be constructed from objects that live naturally on non-flat state-spaces. Our~starting inventory is curves and real-valued functions.  From curves we construct \emph{tangent vectors} and from functions we construct gradient \emph{one-forms}.  Then by taking products we construct \emph{symplectic forms} and \emph{metric forms}; these \emph{bilinear forms} define our state-space geometry.

The classical Hamiltonian framework of Arnol'd \cite{Arnold:199ye} illuminates the dynamical roles of symplectic and metric forms as follows. From the gradient one-form of a \emph{Hamiltonian potential} a \emph{vector field} is constructed that is called the  \emph{Hamiltonian flow}.  Then simulation trajectories are simply the \emph{integral curves} of the flow {(Fig.~\ref{fig: ski}A--B)}.

\subsection{Roles of symplectic flow}
\label{subsec: Roles of symplectic flow} 

The symplectic form respects the first law of thermodynamics (energy conservation) by answering the question,\ ``How~fast and in what direction do we move?'' as follows: ``Given the gradient one-form of a Hamiltonian potential, here is a vector field for which that gradient vanishes; flow tangent to this vector field.''  Figure~\ref{fig: ski}A provides the analogy of skiers traversing a mountainside, with altitude in the role of a Hamiltonian potential.

The second law (nondecreasing entropy) is respected by requiring that the symplectic form be \emph{closed}, so that its \emph{Lie derivative} vanishes, which is the geometric statement of \emph{Liouville's theorem}.  Physically this means that under symplectic flow, state-space density does \emph{not} concentrate.  Figure~\ref{fig: ski}B provides the analogy of a crowd of skiers whose density is unchanging. 

Mathematically speaking, symplectic forms specify a natural \emph{symplectic isomorphism} between one-forms (in this case, the gradient of the Hamiltonian potential) and vector fields (the Hamiltonian flow).

\begin{figure}
\centering%
\vspace{-0.1250\baselineskip}
\includegraphics[width=0.98\columnwidth]{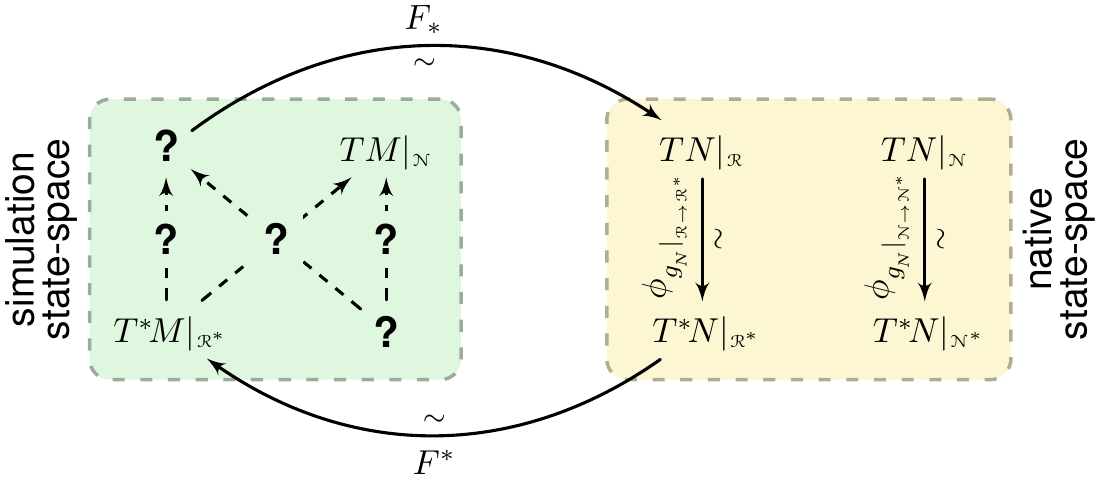}%
\caption[Naturality in projective simulation]{\label{fig: natural}%
We suppose that on a native state-space manifold $N$, the dynamical potentials and the state-space metric $g_N$ have been specified, and we further suppose that the simulation map $F\colon M\to N$ has been specified.  Considering first the native state-space, the following are geometrically  natural constructions for a complete set of \emph{tangent and cotangent subbundles}: 
$T^{\ast\!}N|_{\mathcal{N^ast}} = \Kernel F^\ast$, 
$TN|_{\mathcal{R^ast}} = \Image F_\ast$, 
$T N|_{\mathcal{N}} = \phi^{-1}_{g_N}\left(T^{\ast\!}M|_{\mathcal{N^ast}}\right) = (T N|_{\mathcal{R}})^\perp$, 
and 
$T^{\ast\!}N|_{\mathcal{R}} = \phi_{g_N}\left(T M|_{\mathcal{R}}\right) = (T^{\ast\!}N|_{\mathcal{N}})^\perp$.  
In contrast, on the simulation state-space only two of the four subbundles have natural constructions, namely   
$T^{\ast\!}M|_{\mathcal{R^ast}} = \Image F^\ast$ and 
$T M|_{\mathcal{N^ast}} = \Kernel F_\ast$.   The remaining unspecified elements are labeled \mbox{``{\sffamily\bfseries\upshape?}''}; these must be {designed, validated, and verified} in service of engineering objectives, via the constructions and theorems of Section~\protect\ref{sec: The summit: valid design criteria}.}%
\end{figure}

\subsection{Roles of metric flow}
\label{subsec: Roles of metric flow}   

Now we turn our attention to flows that are metric rather than symplectic. Classical protein simulations often simulate flows induced by interaction with a low-temperature thermal bath, to find (by simulation) protein folds that are low-energy and hence stable.   At~physiological temperatures covalent bonds are effectively rigid links; the conformational state-space then is parameterized by bond angles and has lower dimension than the Cartesian space in which it is immersed.  The pulled-back Cartesian metric endows the bond-angle state-space with a \emph{metric structure}, and by setting the kinetic terms in the Hamiltonian to zero, the state-space is further endowed with a chemical energy function.  

These two new structures---the metric form and the energy function---provide a new answer to the question ``How fast and in what direction do~we~move?'', namely, ``Here is a vector field that maximizes the  energy function's rate-of-change; flow tangent to this vector field'' \mbox{(Fig.~\ref{fig: ski}C)}.   Like symplectic forms, metric forms establish a \emph{metric isomorphism} between one-forms (the energy gradient) and vector fields (the flow of steepest descent).  

We thus arrive at a classical simulation framework whose elements are geometrically natural, in which trajectories are governed by symplectic and metric flows specified in terms of potentials and gradients, and we see how to gain efficiency by adapting state-space geometry to trajectories.  

\subsection{Quantum symplectic flow}
\label{subsec: Quantum symplectic flow}

To extend these principles to quantum systems we first review the symplectic elements of quantum dynamics.  These are well understood \cite{Ashtekar:1999yu,Tyurin:2007dn,Berezin:1975wj}, and we need only specify them in a form that is suited to our geometric simulation framework.

Hamiltonian trajectories---both classical and quantum---are the integral curves of the  Hamiltonian flow $v$ (a vector field) that satisfies  $\phi_\omega(v){\,=\,}\nabla H$, where $H$~is the Hamiltonian potential, $\nabla H\,{\myequiv}\,dH$ is a gradient one-form, $\omega$ is a symplectic form, and (in various notations)\,\ $\phi_\omega(v) \myequiv \omega(v,{\cdot}) \myequiv  i_{v}\omega \myequiv v\,\lrcorner\,\omega$ is the one-form that is induced jointly by $\omega$ and $v$.   

Some authors conventionally define a vector-valued symplectic gradient to be $\nabla_{\omega}\,{\myequiv}\,\phi_{\omega}^{-1}\nabla$, such that Hamilton's equation is $v\,{=}\,\nabla_{\omega}H$; we prefer forms because forms pullback naturally and vectors don't.

To link-up symplectic dynamics with quantum physics, the Schr\"{o}dinger equation $\phi_G(v){\,=\,}-\myJstar\hspace{0.05em}\nabla H/(2\hslash)$ must be satisfied, where $\hslash$~is Planck's constant, $G$~is the \emph{Hilbert space metric}, $\phi_G(v)\myequiv G(v,{\cdot})$ is the metric isomorphism, and $\myJstar$ is a \emph{complex structure} on one-forms (that is, an \emph{automorphism} satisfying $\myJstarinv{\,=\,}\hspace*{-0.05em}-\hspace*{-0.0em}\myJstar$; thus for us the \emph{symplectic gradient} $\myJstar\hspace{0.05em}\nabla$ maps functions $\to$ one-forms).
	
Consistency requires that the Hamilton and Schr\"{o}dinger equations predict the same trajectory.  By direct substitution this requires that the \emph{bijective maps} $\{\phi_G,\phi_\omega,\myJstar\}$ be a \emph{compatible triple} under composition ``$\mycirc$'', \emph{i.e.} $\phi_G\,{=}\,-\myJstar\mycirc\phi_\omega/(2\hslash)$, 
${\phi}_\omega\ {=}\ 2\hslash\,\myJstar\mycirc\phi_G$, and %
	$\myJstar\ {=}\ \phi_\omega\mycirc\phi^{-1}_G/(2\hslash)$.  This triple induces a com\-plex structure on vectors $J\,{\myequiv}\,\phi_G^{-1}\mycirc\myJstar\mycirc\phi_G$ that is a symplectic and metric isometry,  $G(J{\cdot},J{\cdot}){\,=\,}G({\cdot},{\cdot})$ and $\omega(J{\cdot},J{\cdot}){\,=\,}\omega({\cdot},{\cdot})$ (as~is usual in quantum mechanics). 
	
Geometers often further require that $J$ be \emph{integrable}, which we ensure by inducing $J$ via \emph{holomorphic coordinates}; the New\-lander-Nirenberg theorem \cite{Moroianu:07} then guarantees that $J$ is integrable.  The practical advantage is that the natural action of $J$ ($\myJstar$) on the vector (covector) bases that are induced by holomorphic coordinates is simply multiplication by factors of $\pm i$.

The physical lesson is that Schr\"{o}dinger dynamics is symplectic---and thus thermo\-dynamical---if and only if any two of the symplectic, metric, and complex structures $\{\omega, G, J\}$ compatibly determine the third.  A~state-space that is endowed with an integrable compatible triple is, by definition, a \emph{\Kahler\ manifold}\hspace*{0.1em} \cite{Moroianu:07}; specifying quantum dynamics as a symplectic flow on a \Kahler\ manifold is called \emph{geometric quantum mechanics} \cite{Berezin:1975wj,Ashtekar:1999yu,Tyurin:2007dn}; this concludes our summary of it.

\subsection{Quantum metric flow}
\label{subsec: Quantum metric flow}

Symplectic flow alone is insufficient for quantum simulation---just as it is classically insufficient---and so we turn our attention to quantum metric flows.  Metric flows appear naturally in processes involving noise and measurement.  These processes are called \emph{quantum operations}, and their well-known mathematical expression is the \emph{Lindbladian master equation} \cite{Nielsen:00,Sidles:2009cl}.   

A~central theme of our geometric simulation frameworks is that when the Lindbladian equation is \emph{unravelled} \cite{Carmichael:93} in terms of Hamiltonian potentials and their gradients, the resulting flow is concentrative.  

This leads us to envision a quantum simulation framework (Fig.~\ref{fig: ski}C--D) that has the same geometric structure as classical simulation, in which symplectic flow (Schr\"{o}dinger rather than Hamiltonian) unites with metric flow (Lindbladian rather than thermodynamic) and is pulled-back onto a lower-dimension state-space (\Kahlerian\ rather than bond-angle). 

Quantum master equations can be unravelled in various gauge-equivalent ways \cite{Carmichael:93,Nielsen:00}, of which the \emph{synoptic gauge} \cite{Sidles:2009cl} is well-suited to implementing simulations.   This gauge exploits a reciprocal informatic symmetry  \cite{Sidles:2009yq}: ``We~observe a system only by allowing the system to observe us.''  Physically speaking, the synoptic gauge simulates all noise processes as equivalent measurement processes; in effect this describes quantum operations as communication channels.

Our skiing metaphor therefore depicts each Lindblad-\Ito\ skier as descending rather than traversing a mountainside, while observing and responding to the motions of other skiers \dots\ who in turn observe and respond to other skiers {\dots} etc.  We will find that Lindblad-\Ito\ one-forms have a stochastic component that reflects this mutual observation-and-response. 

During the descent we will give detailed account of quantum Lindbladian processes, but first we survey the geometric summit at which we have arrived.

\section{The summit: valid simulation design}
\label{sec: The summit: valid design criteria}

Our ascent has been abstract and rapid---now we survey the summit with a view toward planning our descent to practical applications.  Here we are motivated by Ed Viestur's mountaineering maxim (\onlinecite[][p.~168]{Viesturs:2007lr}) that ``Getting to the top is optional.  Getting down is mandatory.'' 

To assist us with planning our descents to practical applications, we are going to prove three powerful but abstract theorems that relate simulation verification and validation (Tables~\ref{table: metric flow}--\ref{table: symplectic flow}).  We motivate these proofs in the light of two practical examples: geodesic flow on a torus (shown in Fig.~\ref{fig: torus}) and rigid body rotation of a water molecule (shown in Fig.~\ref{fig: rigid}) 

\subsection{Reconciling with Grothendieck}

In discussing our example, we bear in mind a famous passage of Grothendieck that describes his ideal of mathematical naturality:
\begin{quote}
The unknown thing to be known appeared to me as some stretch of earth or hard marl, resisting penetration.\ \ldots\ the sea advances insensibly in silence, nothing seems to happen, nothing moves, the water is so far off you hardly hear it.\ \ldots\  yet it finally surrounds the resistant substance.
\end{quote}
Regrettably, our experience as educators has been that very many students in engineering and science find a Grothendieck-style mode of thinking to be so abstract as to be dauntingly alien; these students learn more easily in the context suggested by Mac Lane's aforementioned maxim ``Mechanics developed by the treatment of many specific problems'' (\onlinecite[][p.~295]{Mac-Lane:1986qe})  and Arnold's aforementioned ``Hamiltonian mechanics is geometry in phase space; phase space has the structure of a symplectic manifold'' (\onlinecite[][p.~161]{Arnold:199ye}).

To reconcile these various styles of naturality, with a view toward making natural simulation frameworks easier to grasp, we are going to motivate our natural-style theorems by Arnold-style and Mac Lane-style consideration of practical examples.  In particular, for us the sea of simulation naturality will not ``advance insensibly in silence'' until it ``surrounds the resistant substance.''  Rather, we will take the point of view that naturality is itself a utilitarian construct of engineering and science, and we will motivate the development of our naturality theorems ``on the fly'' with respect to concrete aspects of our sample problems.

\begin{table}[t]
\vspace{-0.5\baselineskip}
\caption[Compatible metric structure]{\label{table: metric flow}Providing a compatible metric structure to simulation state-space.  See Sec.~\ref{sec: The summit: valid design criteria} for physical motivation and notational details.}
\centering%
\rule[0.35\baselineskip]{\columnwidth}{0.4pt}\newline
{\makebox[\columnwidth][c]{%
\includegraphics[width=\columnwidth]{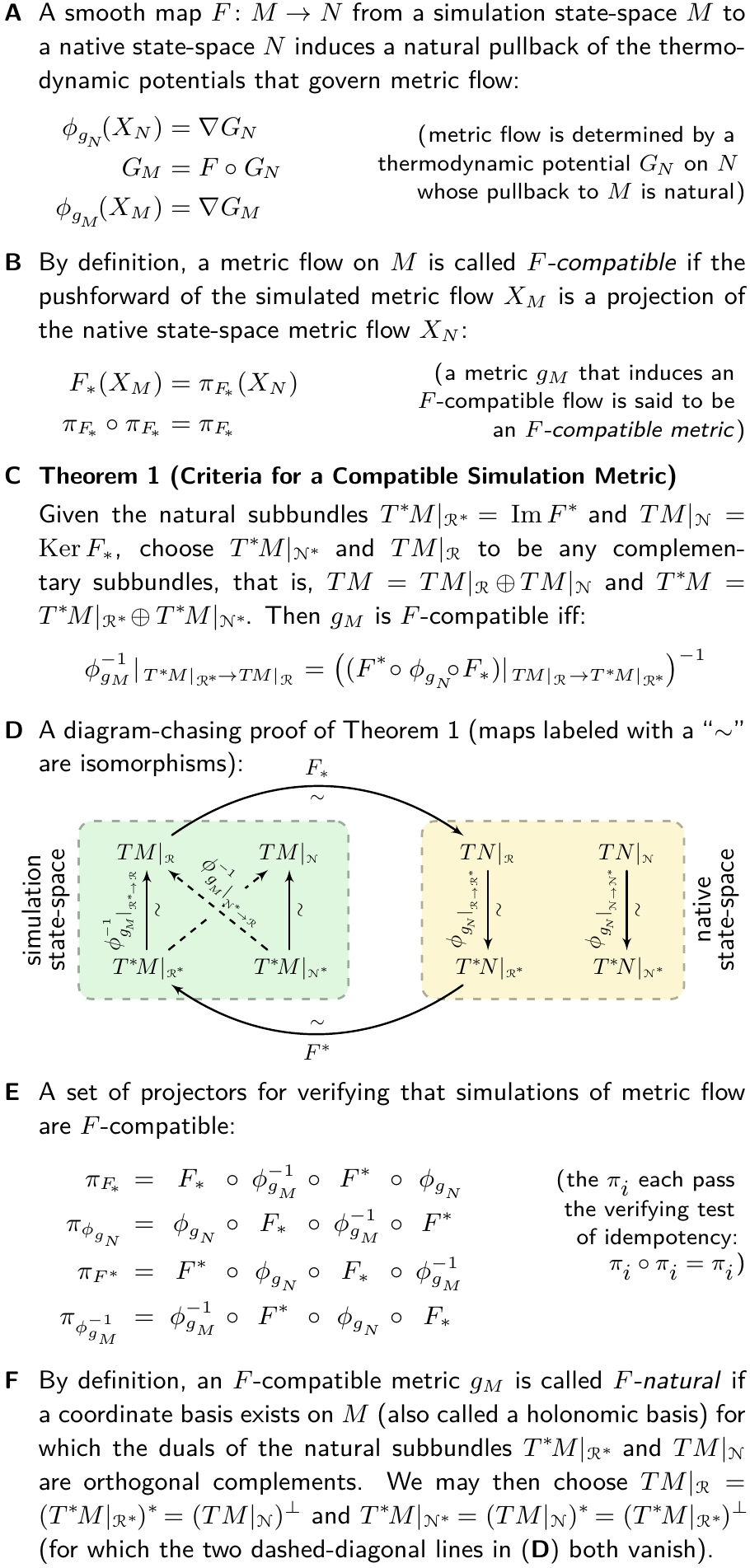}}}%
\newline%
\rule[0.35\baselineskip]{\columnwidth}{0.4pt}
\vspace*{-0.35in}
\end{table}

\subsection{Introducing coordinates}
\label{subsec: introducing coordinates}

We begin with the intuition that the torus is a two-dimensional surface embedded in a three-dimensional Euclidean state-space $N$.  We take Cartesian coordinate functions $N$ to be $\{x^1, x^2, x^3\}$, and we take coordinate functions on the simulation state-space $M$ to be $\{q^0, q^1, q^2, q^3\}$, which are taken to be related to $\{x^1, x^2, x^3\}$ by 
\begin{equation}
		\label{eq: pullback F}
		\left[\begin{array}{c}
		x^1\\
		x^2\\
		x^3
		\end{array}\right] = \left[\begin{array}{l}
		(r_1+r_2\cos \theta)\cos\phi\\
		(r_1+r_2\cos \theta)\sin\phi\\
		r_2\sin \theta
		\end{array}\right]_{\text{\footnotesize\(%
		          \left|\begin{array}{@{\,}r@{}l}
		          \phi&\,=\,\arctantwo(q_0,q_1)\\
		          \theta&\,=\,\arctantwo(q_2,q_3)\,.\hspace{-0.5em}
		          \end{array}\right.\)}
		          }
\end{equation}Already we have departed substantially from traditional descriptions of toric geodesics, which are based upon the above \emph{Clairaut coordinates} $\{\theta, \phi\}$ (see \onlinecite[p.~219 of][]{Oprea:1997uq}), in that we have introduced four coordinate functions $\{q^0, q^1, q^2, q^3\}$ to simulate motion on a two-dimensional manifold.  Thus the map $F: M\to N$ is now specified, but it has \emph{not} been specified in a geometrically natural manner.  Rather, by designating an overcomplete set of coordinates $\{q^0, q^1, q^2, q^3\}$ we will obtain an algebraically efficient simulation.  We will see that choosing coordinates for $F$ is  the \emph{only} step of our simulation framework that is not determined entirely by considerations of naturality.

We further associate to the Euclidean coordinates $\{x^1, x^2, x^3\}$ the momenta $\{p^1, p^2, p^3\}$ and the Hamiltonian function $H=\left((p^1)^2 + (p^2)^2 + (p^3)^2\right)/(2 m)$.  Requiring that the Hamiltonian be a scalar associates to our native state-space a natural metric $g_N: TN\to T^{\ast\!}N$ that evidently is given by \begin{equation}
\label{eq: native metric}
g_N = 2m\left(dx^1\otimes dx^1 + dx^2\otimes dx^2 + dx^3\otimes dx^3\right).
\end{equation}

Here we are including the mass $m$ because when extend this construction to the multi-atom water molecule of Fig.~\ref{fig: rigid}, the individual atoms will have varying masses.

From now on, and in all our examples, we will regard the native metric $g_N$ and the simulation map $F$ to be specified \emph{ab initio}.

\subsection{Constructing natural simulation metrics}
\label{subsec: constructing a natural simulation}

It is evident from the many ``{\sffamily\textbf{?}}'' markings in Fig.~\ref{fig: natural} that the preceding construction is insufficient for specifying a complete set of natural structures on our simulation state-space.  The missing subbundle structures must be \emph{designed, validated, and verified} in service of engineering objectives, and specifying that complete set of natural structures is the objective of (Tables~\ref{table: metric flow}--\ref{table: symplectic flow}). 

We require that our simulations be natural for both metric and symplectic dynamics, and we consider the (simpler) metric case first. which is given in Table~\ref{table: metric flow}. The notation of Table~\ref{table: metric flow} follows without essential change the notation of Lee (\onlinecite{Lee:2003ud}, see pp.~130 and 317--19).  We write \myTableOneEntry{A} to reference the first entry in this table; we adopt this concise notation for all this article's (many) table entries, from \myTableOneEntry{A}--\myTableFourEntry{I}.

Our key engineering objective is that our simulated trajectories be \emph{projectively natural} in the metric sense of \myTableOneEntry{B} and in the symplectic sense of \myTableTwoEntry{B}.

Given a native metric $g_N$ and a simulation map $F$, \myTableOneEntry{F} provides criteria for endowing the simulation state-space with a metric $g_M$ that is \emph{$F$-natural}.   Constructing an $F$-natural $g_M$ is the key step in specifying projectively natural simulations, since all the missing elements labeled ``{\sffamily\textbf{?}}'' in Fig.~\ref{fig: natural} can be filled-in once $g_M$ has been constructed.

The following is a concrete construction of an $F$-natural simulation metric for the torus trajectories of Fig.~\ref{fig: torus}.  The sole inputs to the construction are the simulation map $F$ of (\ref{eq: pullback F}) and the native state-space metric $g_N$ of (\ref{eq: native metric}).  The construction is can be carried through by automated programs \footnote{We used \emph{Mathematica} for all of the (many) symbol manipulations of this article.}. 

\begin{table}[t]
\vspace{-0.5\baselineskip}
\caption[Compatible symplectic structure]{\label{table: symplectic flow}Providing a compatible symplectic structure to simulation state-space. See Sec.~\ref{sec: The summit: valid design criteria} for physical motivation and notational details.}
\centering%
\rule[0.35\baselineskip]{\columnwidth}{0.4pt}\newline
{\makebox[\columnwidth][c]{%
\includegraphics[width=\columnwidth]{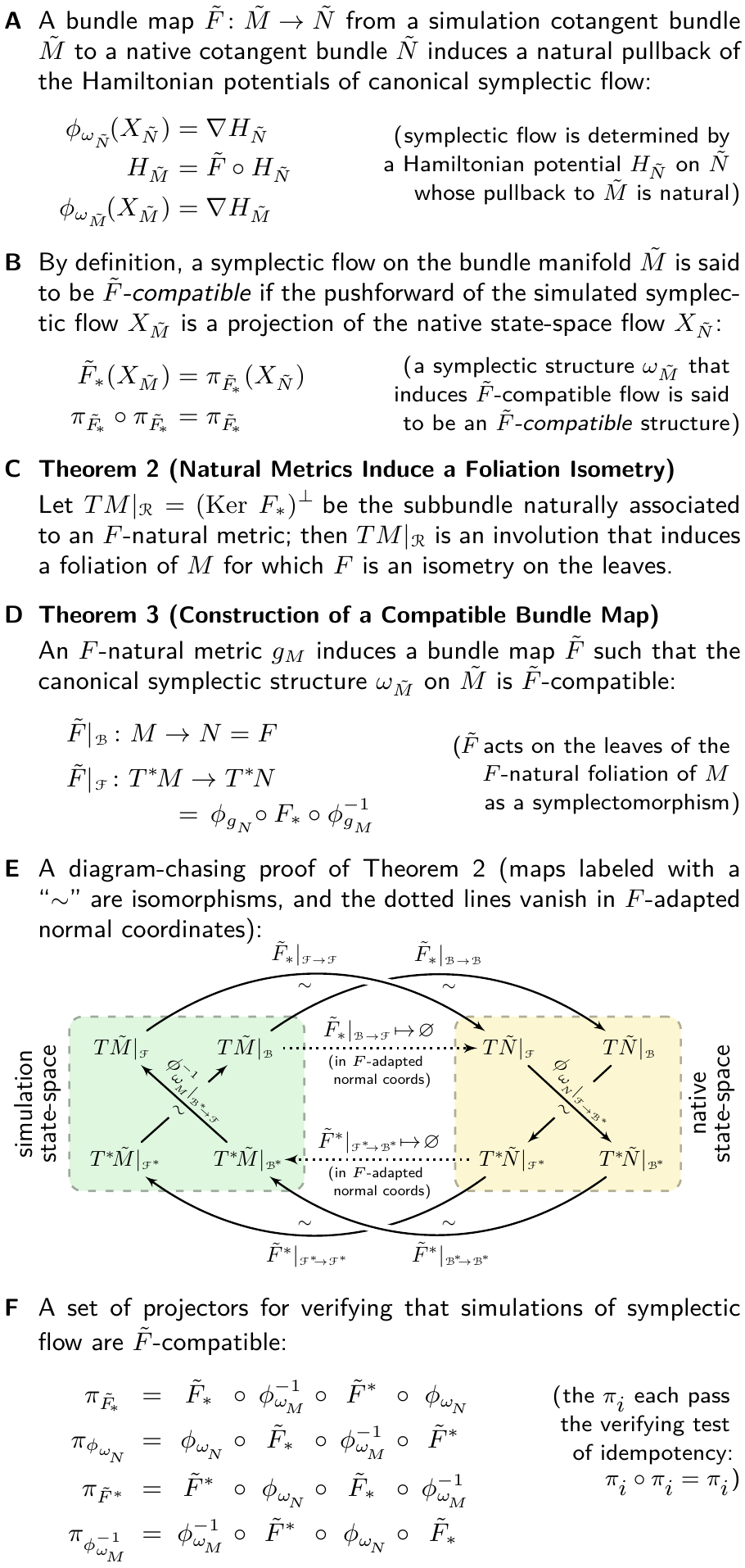}}}%
\newline%
\rule[0.35\baselineskip]{\columnwidth}{0.4pt}
\vspace*{-0.35in}
\end{table}

In accord with \myTableOneEntry{B} the pulled-back native metric $F^\ast\circ g_N\circ F_\ast$ specifies two natural subbundles on $M$, namely $T^{\ast\!}M|_{\mathcal{R^{\ast\!}}}$ and $TM|_{\mathcal{N}}$.  Although the subbundles themselves are natural objects, there is in general no unique natural choice of basis vectors for them.  By a singular value decomposition (\SVD) of the matrix representation of $F^\ast\circ g_N\circ F_\ast$ in the chosen coordinate basis, we may choose the subbundles to be spanned by the \SVD\ basis vectors{:} \begin{align}
\label{eq: first basis}
T^{\ast\!}M|_{\mathcal{R^{\ast\!}}} & = \Span 
\begin{alignedat}[t]{3}
(\,&&
	q^1 dq^0&- q^0 dq^1&&,
\\[-0.25ex]
&&
	q^3 dq^2&- q^2 dq^3
&&\,)
\end{alignedat}
\\[0.75ex]
\label{eq: second basis}
TM|_{\mathcal{N}} &= \Span 
\begin{alignedat}[t]{3}
(\,&&
	q^0\partial_{q^0}&+q^1\partial_{q^1}
&&,\\[-0.25ex]&&
	q^2\partial_{q^2}&+q^3\partial_{q^3}
&&\,)
\end{alignedat}
\end{align}
Now comes the sole step in the construction that is not geometrically natural: the complementary subbundles $TM|_{\mathcal{R^{\star\!}}}$ and $T^{\star\!}M|_{\mathcal{N^{\star\!}}}$ are specified as coordinate duals: $TM|_{\mathcal{R}} = (T^{\ast\!}M|_{\mathcal{R^{\ast\!}}})^{\ast\!}$ and  $T^{\ast\!}M|_{\mathcal{N^{\ast\!}}} = (TM|_{\mathcal{N}})^{\ast\!}$.  The subbundles so constructed then satisfy
\begin{align}
\epsilon(E) & =0\text{\ \,for\ \,} \epsilon \in T^{\ast\!}M|_{\mathcal{R^{\ast\!}}}\,,\,E \in TM|_{\mathcal{N}},\ \text{and}\\
\epsilon(E) & =0\text{\ \,for\ \,} \epsilon \in T^{\ast\!}M|_{\mathcal{N^{\ast\!}}}\,,\,E \in TM|_{\mathcal{R}}
\end{align}
Note that the above subbundle relations are geometrically natural, and thus hold in all coordinate systems, even though they refer to subbundles that were constructed by \SVD\ methods in a particular coordinate system associated with (\ref{eq: pullback F}--\ref{eq: second basis}).  Because these relations hold in all coordinate systems, the remainder of our analysis can be carried through entirely via natural constructions.

Following the prescription of \myTableOneEntry{C} we then construct an $F$-natural metric $g_M$ having the property that the above subbundles are simultaneously coordinate duals and orthogonal complements, as specified in \myTableOneEntry{F}.  Such metrics are said to be \emph{$F$-natural} because, if we construct (suitably) \emph{Riemannian normal coordinates} on $M$ and $N$, then it is straightforward to verify that the four bundle maps $\{\phi_{g_N}, \phi_{g_M}, F_\ast, F^{\ast}\}$ can be made locally diagonal by a suitable adaptation of these coordinates.   We will not give the details of the normal coordinate construction, which is straightforward but rather lengthy; the required computational idioms are thoroughly covered in Lee's two books (\onlinecite[see Thms.~7.8--13, 11.24, 19.10, and A.33 of ref{.}][]{Lee:2003ud}, or alternatively \onlinecite[see Props.~5.7 to 5.11 of ref{.}][]{Lee:1997uq}).

In the adapted normal coordinates so constructed, Theorems~2 and~3 of Table~\ref{table: symplectic flow} are trivially true, and because the theorems are covariantly stated, they hold in all coordinate systems; similarly the verification operators $\pi_i$ are \myTableTwoEntry{E} projectors relative to any basis.  Thus the main utility of normal coordinates is theorem-proving; in integrating dynamics trajectories we need not compute them.

\subsection{Dimension reduction and augmentation}

In essence what the theorems and constructions of Tables~\ref{table: metric flow}--\ref{table: metric flow} provide is a general method, not only for dimension reduction, but also for dimension augmentation.  The theorems and constructions  are particularly useful in cases in which dimension augmentation allows dynamical equations to be cast into a computationally advantageous algebraic form.  

As we will discuss later on, in the context of quantum simulation, Hilbert space can be be viewed as a dimensional augmentation of the (unknown) underlying quantum state-space of nature; thus dimension augmentation methods are of considerable fundamental interest in themselves, in addition to their practical utility in simulations.

\subsection{Rigid body rotation}
\label{subsec: rigid body rotation}

Now let us construct a dynamical simulation of rigid body motion by the methods of the preceding section.   Our example will be the tumbling water molecule of Fig.~\ref{fig: rigid}, and the symplectic dynamical equations that we derive will be of a quaternionic variety has found much use in astronomy \cite{Cid:1987fk} and in molecular simulation \cite{Miller-III:2002fk}.  Our main goal here is to show that once quaternionic coordinates have been specified, then the derivation of the equations can be carried through automatically, via Theorems 1--3 of Tables~\ref{table: metric flow}--\ref{table: symplectic flow} and their associated natural constructions.

Let Euclidean coordinates on the native state-space be a set of triplets $\{(x_i^1, x^2_i, x^3_i)\}; i \in (1,\dots, n_\text{atom})$ associated to canonical momenta $\{(p_i^1, p^2_i, p^3_i)\}; i \in (1,\dots, n_\text{atom})$.  Let the Hamiltonian potential be
\begin{equation}
	H = \sum_{i = 1}^{n_\text{atom}}\,\left((p^1_i)^2 + (p^2_i)^2 + (p^3_i)^2\right)/(2 m_i)
\end{equation}
where $m_i$ is the mass of the $i$-th atom of the rigid body.  We require that this potential be a geometric scalar, and from this requirement we read-off the dynamical metric $g_N$ of the native atomic state-space to be:
\begin{equation}
g_N = \sum_{i = 1}^{n_\text{atom}} \sum_{j = 1}^{3}\,(2 m_i) \, (dx^j_i \otimes dx^j_i)
\end{equation}
We specify the coordinate map $F$ by
	\begin{equation}
			\left[\begin{array}{@{\,}c@{\,}}
			(x^1)_i\\[0.5ex]
			(x^2)_i\\[0.5ex]
			(x^3)_i
			\end{array}\right] = 
			\left[\begin{array}{@{\,}c@{\,}}
			Q^1\\[0.5ex]
			Q^2\\[0.5ex]
			Q^3
			\end{array}\right] 		
	+
			\left[\begin{array}{@{\,}c@{\,}}
			R(q^0\!,q^1\!,q^2\!,q^3)\\
			\end{array}\right]
			\left[\begin{array}{@{\,}c@{\,}}
			(x^1_0)_i\\[0.5ex]
			(x^2_0)_i\\[0.5ex]
			(x^3_0)_i
			\end{array}\right] 		
	\end{equation}
where by inspection $(Q_1,Q_2,Q_3)$ are center-of-mass coordinates and $R(q^0\!,q^1\!,q^2\!,q^3)$ is a $3\,\times\,3$ rotation matrix that is a function of quaternionic coordinates $\{q^0\!,q^1\!,q^2\!,q^3\}$.  

A quick way to derive the quaternionic functional form of $R(q^0\!,q^1\!,q^2\!,q^3)$ is via Pauli spin matrices $\{\sigma^0\!,\sigma^1\!,\sigma^2\!,\sigma^3\}$ in which case the rotational transform $R(\lb{q}): \lbhat{x}\to\lbhat{x}'$ is given by
	\begin{align}
	\label{eq: pauli one}
	(\hat{\mathbf{x}}'\cdot\boldsymbol{\sigma}) & = 
	R^\dagger(\mathbf{q})(\hat{\mathbf{x}}\cdot \boldsymbol{\sigma})R(\mathbf{q})\\
	\intertext{where}
	\label{eq: pauli two}
	R(\boldsymbol{q}) & = \frac{q^0\sigma^0 - iq^1\sigma^1 - iq^2\sigma^2 - iq^3\sigma^3}
	{\big(\,(q^0)^2 +(q^1)^2 +(q^2)^2 +(q^3)^2\,\big). }
	\end{align}
We recover the standard quaternionic representation for $R$ acting on the unit vector $\lbhat{x}$ by taking traces of (\ref{eq: pauli one}) with (\ref{eq: pauli two}) substituted.
	
Then just as we calculated the two natural toroidal subbundles $T^{\ast\!}M|_\mathcal{R^{\ast\!}}$ and $TM|_\mathcal{N}$ of (\ref{eq: first basis}--\ref{eq: second basis}), we calculate the two natural rigid body subbundles to be 
	\begin{align}
	T^{\ast\!}M|_\mathcal{R^{\ast\!}} & = 
	\begin{alignedat}[t]{8}
		&\makebox[0pt][l]{$\Span(\,dQ^1, dQ^2, dQ^3,\,$}&\\
		&\,&-\,q^1\,dq^0
		&\,&+\,q^0\,dq^1
		&\,&-\,q^3\,dq^2
		&\,&+\,q^2\,dq^3,
	&&\\
		&\,&-\,q^2\,dq^0
		&\,&+\,q^3\,dq^1
		&\,&+\,q^0\,dq^2
		&\,&-\,q^1\,dq^3,
	&&\\
		&\,&-\,q^3\,dq^0
		&\,&-\,q^2\,dq^1
		&\,&+\,q^1\,dq^2
		&\,&+\,q^0\,dq^3)
	&&
	\end{alignedat}\hspace{-2em}\\[0.75ex]
	TM|_\mathcal{N}&= \Span 
	\begin{alignedat}[t]{1}
	(&q^0\partial_{q^0}
		&&\,{+}\,q^1\partial_{q^1}
		&&\,{+}\,q^2\partial_{q^2}
		&&\,{+}\,q^3\partial_{q^3}
	&&)
	\end{alignedat}
	\end{align}
The simple algebraic form of these quaternionic subbundles is, from a natural point of view, the sole (and sufficient) justification for embracing quaternionic coordinates.  

The remainder of the rigid body dynamical calculations shown in Fig.~\ref{fig: rigid} proceeds according to the theorems and constructions of Tables~\ref{table: metric flow}--\ref{table: symplectic flow}, precisely as for the previous case of geodesic motion on a torus.  We will not give details, because the successive steps each are geometrically natural and the expressions obtained are equivalent to (for example) the \NOSQUISH\ formalism \cite{Miller-III:2002fk} of the molecular dynamics literature.  

The chief practical advantage of deriving the rigid-body dynamical equations by this route is conceptual simplicity: Theorems 1--3 guarantee that the simulation dynamics are symplectic, the proofs of the theorems are diagrammatic, the associated constructions are generically natural and thus can be automated, and the concrete validation and verification criteria of \myTableOneEntry{F} can be used to check the resulting dynamical simulation codes.

It is no use pretending, though, that natural simulation methods inevitably produce dynamical simulation codes that are particularly short or simple.  Even for our simple example of a water molecule, the projector $\pi_{\omega_N}$ of \myTableOneEntry{F} is an $18\,\times\,18$ matrix whose associated projective verification $\pi_{\omega_{N\!}}\cdot\pi_{\omega_{N\!}} = \pi_{\omega_N}$ requires the separate cancellation of $18^2\,{=}\,324$ matrix entries.   

The simulation of rigid body motion in quaternionic coordinates thus marks a level of software complexity at which the availability of concrete methods for validation and verification starts to become a practical necessity. 

\section{The descent: quantum simulation}
\label{sec: The descent: practical quantum}

Now we have gathered all of the physical ideas and mathematical tools needed to construct a quantum simulation framework in three explicit stages:  first we specify orthodox quantum dynamics in terms of forms on Hilbert state-spaces; then we pull-back the forms onto low-dimension \Kahlerian\ state-spaces; and finally we specify efficient algorithms for integrating the flow of trajectories.   In detail, this construction is accomplished by the chain of reasoning given in Table~3, whose geometric and algebraic essence is the natural pullback of potentials and forms and the natural \emph{pushforward} of trajectory curves; thus the individual steps of Table~3 require only elementary manipulations that can be largely automated.    

\begin{table*}
\vspace*{-0.5\baselineskip}%
\caption[Simulating quantum dynamics]{\label{table: summary}Simulating metric and symplectic quantum dynamics on \Kahlerian\  state-spaces.}%
\centering%
\scalebox{1.0}[1.0]{\makebox[\textwidth][c]{%
\includegraphics[width=1.0125\textwidth]{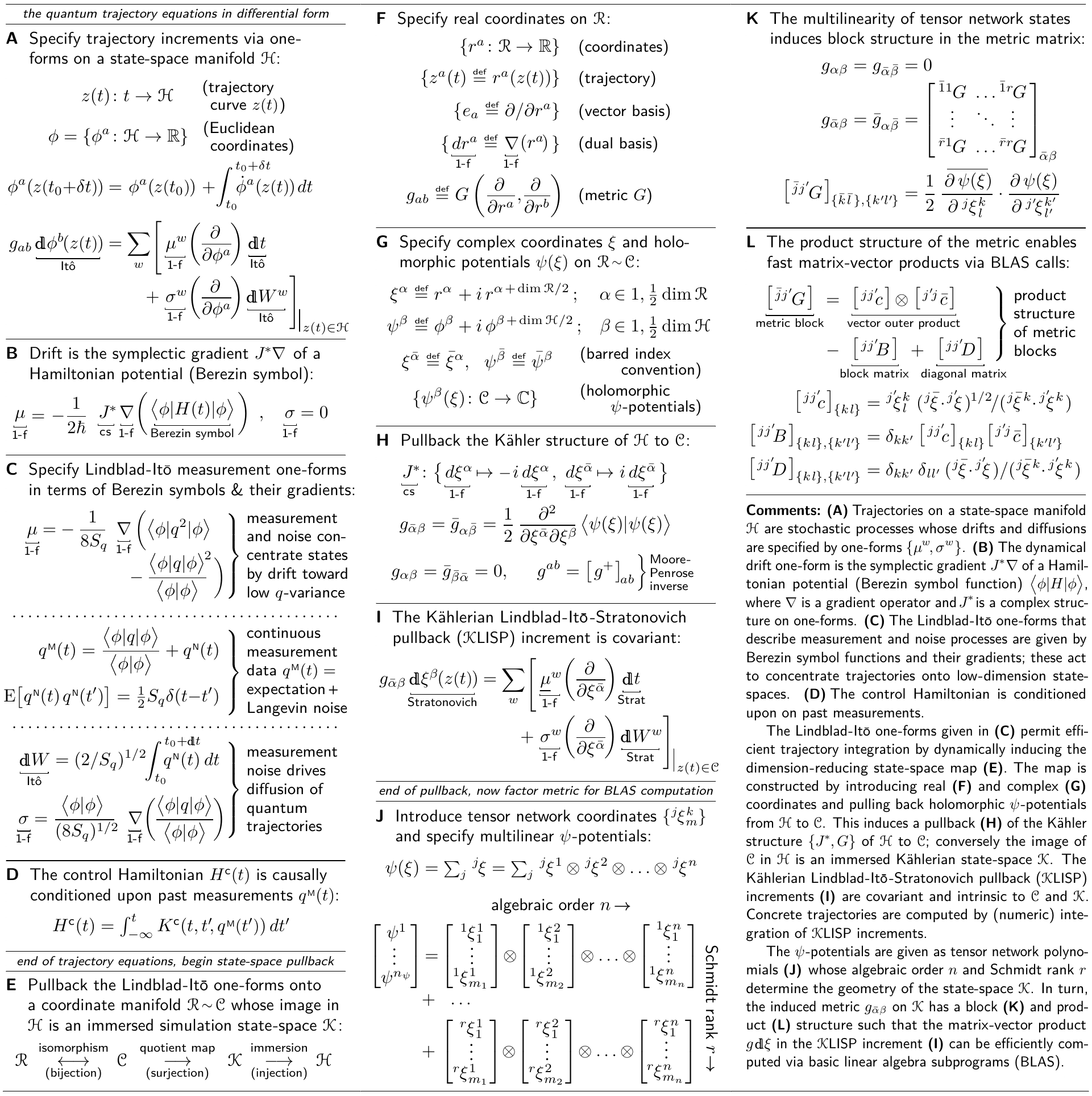}}}%
\vspace{-1.25\baselineskip}%
\end{table*}

\begin{figure*}
{\centering%
\vspace*{-1.5\baselineskip}%
\includegraphics[width=0.95\textwidth]{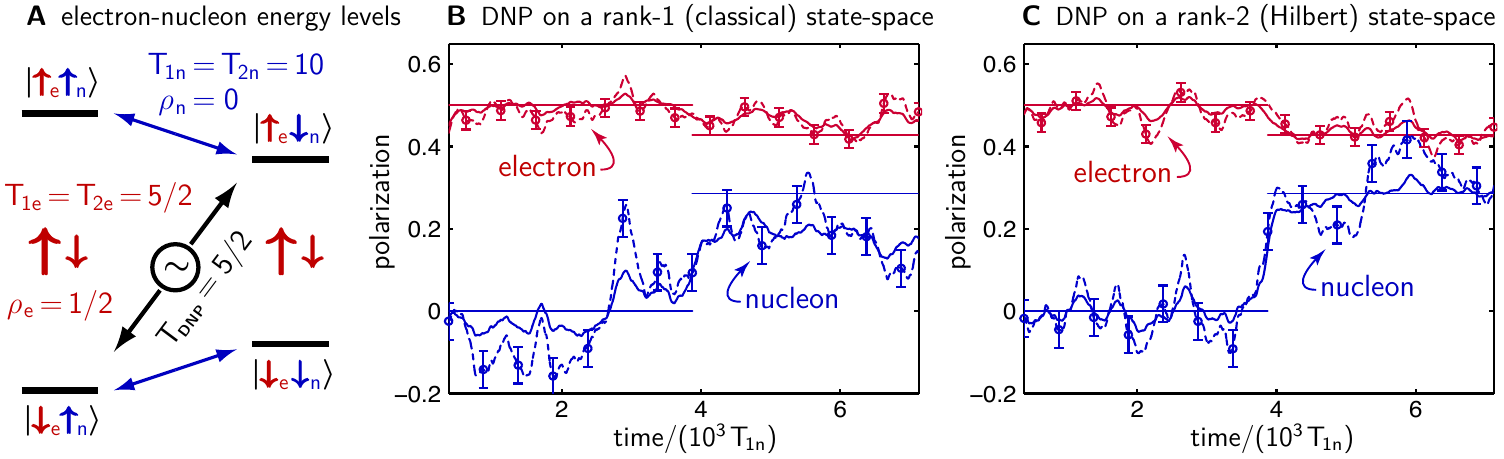}%
\caption[Dynamic nuclear polarization (\DNP)]{\label{fig: DNP}%
{%
%
{\sffamily\mdseries Pullback simulation of dynamic nuclear polarization (\DNP).}\hspace{0.35em}%
{\sffamily\bfseries (A)\,}%
The relaxation time constants and equil\-ibrium polarization values of an electron-nucleon system  are given in terms of Bloch parameters 
	${\{T_{1\text{e}}, T_{2\text{e}}, \rho{_\text{e}}}\}$ 
and 
	${\{T_{1\text{n}}, T_{2\text{n}}, \rho_{\text{n}}}\}$ 
(numerical values are given in the figure). %
Switching on the dynamic polarization process at time $t\,{=}\,4$ alters the (analytic) mean polarization from  
	${\rho{_\text{e}}\,{=}\,{1/2}}$, 
	${\rho{_\text{n}}\,{=}\,{0}}$ 
to 
	${\rho{_\text{e}}\,{=}\,{3/7}}$, 
	${\rho{_\text{n}}\,{=}\,{2/7}}$ 
(analytic values are drawn as straight lines).  
\quad{\sffamily\bfseries (B)\,}%
Trajectories integrated on a rank-1  tensor network state-space have the same dynamical degrees of freedom as a classical two-spin Bloch state-space.    Solid lines show the fluctuating polarization of the simulated quantum state trajectory; dashed lines show the thermal reservoir's (equivalent) continuous measurement of the polarization.  Measurement data are low-pass filtered at 6~dB per octave with filter poles at 
	${{1/(125}\,T_{1\text{n}}{)}}$%
\ with error bars drawn at 
	${\pm1}$ 
standard deviation of filtered continuous polarization estimate.  
The simulated mean dynamic nuclear polarization is 
	${\rho_{\text{n}}\simeq{0.18}}$, 
which is 
	${{\sim}\,{63}\%}$ 
of the Hilbert-space value.%
\quad{\sffamily\bfseries (C)\,}
Trajectories integrated on a rank-2 tensor network state-space have the same dynamical degrees of freedom as quantum spins on a Hilbert space. The polarization estimates accord with the analytic Hilbert-space values within $\pm0.02$.  In trajectory simulations as in experiments, lengthy integration time s  ($7000\,T_{1\text{n}}$ in the present case) are required to achieve reasonable statistical accuracy.
}%
}}%
\end{figure*}

\subsection{Trajectories are integral curves}
We first specify the simulation trajectory as an integral curve on a metric state-space \mbox{(Table~3A;} henceforth {``{\myTableThreeEntry{A}}'', {etc{.}}\ for concision)}.  The trajectory increments {\myTableThreeEntry{A}}~are specified in terms of one-forms multiplied by \emph{\Ito\ increments} \cite{Kloeden:1992kx} $\mydelta t$ and $\mydelta W$.  Note that stochastic increments are expressed by a double-struck ``$\mydelta$''; we reserve the single-struck  ``$d$'' for \emph{exterior derivatives} \cite{Moroianu:07,Arnold:199ye}.  

In {\myTableThreeEntry{B}} we discern the dynamical one-form of the Schr\"{o}dinger equation previously described.  The familiar \emph{Dirac braket} of quantum mechanics is here regarded as a real-valued function on a general \Kahlerian\ state-space; we follow Tyurin \cite{Tyurin:2007dn} in calling these functions \emph{Berezin symbols} \cite{Berezin:1975wj}, although we have retained the physics bra-ket notation in token of its familiarity.  The symplectic gradient $\myJstar\nabla$ enters in the Schr\"{o}dinger dynamics in~accord with the traverse skiing metaphor of Fig.~\ref{fig: ski}A--B. 

\subsection{Lindblad-\Ito\ one-forms}

In \myTableThreeEntry{C} we give the Lindblad-\Ito\ \cite{Adler:00,Caves:00,Nielsen:00} one-forms associated with measurement of a general operator $q$, which are seen to have both a \emph{drift component} and a \emph{diffusion component}.  These one-forms are given in the synoptic \cite{Sidles:2009cl} Lindbladian gauge that concentrates trajectories.  We see that only the metric-flow gradient $\nabla$ enters in the one-forms---not the symplectic-flow gradient $\myJstar\nabla$---in~accord with the downhill skiing metaphor of Fig.~\ref{fig: ski}C--D. 

The Lindblad-\Ito\ dynamical forms of \myTableThreeEntry{C} supply us also with a continuous measurement of $q^{\scriptscriptstyle\text{M}}(t)$; this provides causal Hamiltonian control via {\myTableThreeEntry{D}}.   

In the appended material the dynamical forms {\myTableThreeEntry{A--D}} are elaborated into recipes for simulating arbitrary Lindbladian processes (a detailed derivation is given).  The resulting dynamical forms are simple variations on {\myTableThreeEntry{A--D}}, with \emph{Bloch relaxation} as a special case.  Thus {\myTableThreeEntry{A--D}} comprise a complete set of quantum postulates---equivalent to Postulates 1--4 of Nielsen and Chuang \cite{Nielsen:00}, for example---given in terms of geometric forms and dynamical flow.  

\subsection{Lindblad reduction}

That the dynamics of {\myTableThreeEntry{A--D}} acts to concentrate trajectories onto lower-dimension manifolds has long been noted in the fundamental physics literature.  Adler and Bassi \cite{Adler:2009vh} have reviewed quantum theories that (in effect) concentrate trajectories by gravitationally measuring a localized Hamiltonian~$H_\text{\bfseries local}$.  Especially, Hughston has postulated a stochastic state-reduction process \cite[Eq.~8.1 of][]{Hughston:1996eu} that is equivalent to the special case of \myTableThreeEntry{A--C} with $q\,{=}\,H_\text{\rmfamily local}$ and %
$S_q\,{=}\,(\hbar^3 c^5/G_{\scriptscriptstyle\text{\rmfamily N}})^{1/2\!}/8$,\ where $G_{\scriptscriptstyle\text{\rmfamily N}}$ is the gravitational constant. %
Our simulation framework exploits this natural concentrative dynamics for practical purposes.

Thus, ideas from fundamental physics furnish the middle panel of a compatible triptych of motivations for studying the concentrative one-forms {\myTableThreeEntry{A--D}}: geometers conceive them as natural; physicists postulate them as fundamental; engineers apply them as agents of concentration.

\subsection{The roles of measurement}

Already we can discern the general form of concen\-tration-and-pull\-back simulations, and we can inspect the dynamic nuclear polarization (\DNP) simulation of Fig.~\ref{fig: DNP} with good understanding.  The output of the simulation is a stream of measurements---exactly as in real-world experiments.  Questions about measurement statistics also are well posed, for example: ``What is the mean value of $q^{\scriptscriptstyle\text{M}}(t)$?''  These capabilities allow us to duplicate, by averaging trajectories and measurements, predictions that in Hilbert-space simulation frameworks are achieved by \emph{density matrix} methods.

Physically speaking, in the \DNP simulations of Fig.~\ref{fig: DNP}, environmental processes continuously measure and control the spin direction, inducing relaxation in accord with specified Bloch parameters.  High-order quantum correlations that otherwise would be costly to simulate are quenched by the dynamical flow of trajectories onto low-dimension submanifolds.

A variety of concentration theorems follow from the dynamical one-forms of {\myTableThreeEntry{A--D}} \cite{Sidles:2009cl}.   In particular, trajectories can be thermalized and concentrated onto a lower-dimension product-space of \emph{coherent states} by measuring and controlling all three spin axes simultaneously, such that the pulled-back \emph{Fokker-Planck equations} yield \emph{positive $P$-representations} of thermal density matrices \cite{Sidles:2009cl}.  Physically, these theorems specify the dynamical one-forms of concentration and thermalization.

\subsection{Quantum trajectory pullback}

Our next task is to pullback the dynamical one-forms of  {\myTableThreeEntry{A--D}} onto a reduced-dimension state-space $\lcal{K}$ that is holomorphically immersed in the Hilbert space $\lcal{H}$, so that $\lcal{K}$ inherits a \Kahler\ structure by pullback.  To accomplish this, in {\myTableThreeEntry{E--H}} we introduce a real coordinate manifold $\mathcal{R}$ that is isomorphic to a complex coordinate manifold $\mathcal{C}$  and we equip $\mathcal{C}$ with a set of holomorphic functions $\{\psi\}$.  We identify the $\psi$-functions with (canonical) Hilbert coordinates on \lcal{H}; this induces a natural immersion in $\lcal{H}$ of the \emph{quotient manifold} $\lcal{K}\myequiv\lcal{C}/\{\psi\}$; thus a set of $\psi$-functions concretely specifies a \Kahler\ geometry for the state-space $\lcal{K}$.  Physically we choose geometries that match the concentrative dynamics of the systems we are simulating.

In {\myTableThreeEntry{I}} the Hilbert one-forms of {\myTableThreeEntry{A}} are specified in \emph{Stratonovich form}; these one-forms pullback covariantly \cite{Kloeden:1992kx} and so are geometrically natural.   The sup\-port\-ing material provides a com\-pen\-dium of \Kahlerian-Lind\-blad-\Ito-Stra\-tono\-vich pull\-back (\KLISP ) one-forms that is Lind\-blad-complete, that is, sufficient to simulate the trajectories of any Lindbladian process \mbox{(including} the Bloch-type dynamic polarization process that is simulated in Fig.~\ref{fig: DNP}). 

\subsection{Tensor-network state-spaces}

Now we are ready to choose a class of geometries for $\lcal{K}$, upon which we will compute the simulations of Fig{.}~\ref{fig: DNP}.  We choose the \emph{tensor network manifolds} of {\myTableThreeEntry{J}}.  These ubiquitous \cite{Loan:2000hb} state-spaces are ruled \cite{Sidles:2009cl} hyperbolic \cite{Taimina:2009sy} manifolds; physicists call them \emph{matrix product states} (\MPS).  The states   {\myTableThreeEntry{J}} are a particular class of \MPS whose matrices are diagonal \cite[see Fig{.}~3 of][]{Sidles:2009cl}\hspace*{0.05em}, so that $r$ is their \emph{Schmidt rank}.   Because multiplication of diagonal matrices is Abelian, these states are natural for simulating non-ordered systems. Geometrically the states {\myTableThreeEntry{J}} are a \emph{join space} \cite{Prenowitz:1979by} of multilinear \emph{algebraic varieties} whose $2^{n-\scalebox{0.9}{\mbox{$\scriptstyle\hspace*{-0.1em}1\!$}}}$ bipartitions all have Schmidt rank~${\le}r$.  

This versatile class of manifolds readily 
accommodates even (non-ordered) classical dynamics, because the pullback of the Hilbert space symplectic form $\omega$ onto the Schmidt rank-1 product of normalized coherent states $\otimes^{\rule[-0.4ex]{0pt}{0ex}\smash{\!n_{\scriptscriptstyle\text{\bfseries nspin}}}}_{k{=}1}\,\ket{j_{k},\lbhat{x}_k}$ is simply\[%
\omega_{\scriptstyle\text{Bloch\!\!}}\, = \sum_{k{=}1}^{\rule[-0.5ex]{0pt}{1.35ex}{n_{\scriptscriptstyle\text{spin\!}}}}\ \hbar\ j_k\ \epsilon_{abc}\ x^{a}_k\,[dx^{b}_k \otimes dx^{c}_{\rule[-0.40ex]{0pt}{1ex}k}],\] %
where $j_{k\!}\in\{\tfrac{1}{2},1,\tfrac{3}{2},\ldots\}$ is the $k$'th spin's quantum number and $\lbhat{x}_{k}=\{x^1_k,x^2_k,x^3_k\}$ is a unit vector on its \emph{Bloch sphere}; this is the symplectic form of classical spins \cite{Gustavsson:2009qt} and (as~$j_{k\!}\,{\to}\,\infty$) of classical test-masses \cite{Sidles:2009cl}.

A key state-space parameter is the Schmidt rank~$r$; as $r$ increases (resp.\ decreases), the flow becomes more quantum (resp.\ more classical).
\mbox{Example:} the $r\,{=}\,1$ state-space of Fig.~\ref{fig: DNP}B has dimensionality  $%
\mydim_{\mathbb{R}}\lcal{K}%
	\ {\myequiv}\ %
\rank\raisebox{-0.1ex}[\height][0pt]{$[$}%
g\raisebox{-0.1ex}[\height][0pt]{$]$}%
	\ {=}\ %
6$; physically these six~dimensions are the two Bloch angles of each of $n\,{=}\,2$ spins plus the two dimensions of a (dynamically irrelevant) complex amplitude; the resultant dynamics is exactly classical.  Upon increasing the Schmidt rank to $r\,{=}\,2$ in Fig.~\ref{fig: DNP}C we have $\mydim_{\mathbb{R}}\lcal{K}%
\ {=}\ 8\ {=}\  2{\,\times\,}2^n\ {=}\ \mydim_{\mathbb{R}}\lcal{H}$; the two added dimensions encode \emph{entanglement} and the dynamics is fully quantum.  Thus tensor networks can be tuned to carry any required level of quantum entanglement by adapting the Schmidt rank.  

\subsection{Factored metric representations}

To illustrate the computational challenges of state-spaces having dimension intermediate between classical and Hilbert, we consider a tensor network having order (say) $n\,{=}\,500$ spins and Schmidt rank (say) $r\,{=}\,100$.  This bench\-mark is selected with a view toward simulations in imaging and spectroscopy: there are $n\,{=}\,500$ proton spins in $8~\text{nm}^{3\!}$ of water, which is comparable to present-day imaging voxels \cite{Sidles:2009yq}\!, and the Schmidt rank $r\,{=}\,100$ encompasses $2r\,{=}\,200$ degrees of freedom per spin; thus the state-space has $100{\times}$ the dimensions of classical Bloch dynamics, but only an infinitesimal portion of the $\mydim_{\mathbb{R}}\lcal{H}\,{=}\,4^{n\!}\,{=}\,2^{1000}$ Hilbert dimensions. %

Computing the trajectory increment~$\mydelta\xi$ amounts to solving the \KLISP equations {\myTableThreeEntry{I}} by computing the isomorphism $\phi^{-1}_G$ (called by physicists the \emph{raising and lowering indices}, variously called by mathematicians the  {canonical}, {metric}, {natural}, or musical isomorphism) that maps one-forms onto tangent vectors.  A~key task in iterative algorithms for solving matrix equations \cite{Greenbaum:1997nr} is computing the matrix-vector product $g\,\mydelta\xi$, where $g$ is a $2nr\,{\times}\,2nr$ Hermitian matrix; thus naively evaluating $g\,\mydelta\xi$ for $n\,{=}\,500$ and $r\,{=}\,100$  requires {$4r^2n^2\,{=}\,10^{10}$} multiplications and storage elements.

Seeking efficiency in evaluating $g\,\mydelta\xi$, we find that tensor network structure induces a metric block structure {\myTableThreeEntry{K}} such that 
evaluating $g\,\mydelta\xi$ via {\myTableThreeEntry{L}} requires only \smash{$8r^2n=4{\cdot}10^7$} multiplications and \smash{$6r^2n=3{\cdot}10^7$} storage elements.   The resultant $n/2$ acceleration and $2n/3$ memory compression allows the $g\,\mydelta\xi$ product for $n\,{=}\,500$, $r\,{=}\,100$ to be feasibly computed even on small computers\footnote{The matrix-vector product for $n\,{=}\,500$, $r\,{=}\,100$ requires 0.88 seconds on an {A}pple {M}ac{B}ook laptop.}. 

As with Navier-Stokes simulations, the availability of $\mathcal{O}(n)$ algorithms for matrix-vector products greatly extends the range of practical simulation applications.

\subsection{Physical significance of Schmidt rank}

We have seen that noise levels and Schmidt rank play roles in our simulation framework that are broadly analogous to artificial viscosity and finite-element size in Navier-Stokes simulations, that~is, they control the trade-off of computation cost against simulation accuracy.   A strong concentration conjecture is that spatially extended, dipole-coupled, high-temperature spin systems---such as occur in imaging and spectroscopy---are dynamically concentrated onto state-spaces of bounded Schmidt rank $r$; this conjecture implies that simulating these spin systems is feasible with classical resources.  Numerical investigations extending to $n\,{=}\,18$ quantum spins on tensor network spaces of Schmidt rank up to $r\,{=}\,50$ provide some evidence for this conjecture \cite[see Figs.\ 11--13 of][]{Sidles:2009cl}. 

If further investigations continue to reveal stronger concentration theorems and better numerical algorithms---such that pullback remediates (in part) the curse of dimensionality---then eventually quantum simulation capabilities may parallel those of Navier-Stokes simulation.

\section{Five natural simulation challenges}
\label{sec: Five natural challenges}

Now we will suggest five challenges that arise in the context of natural simulation.   This enumeration is not intended to be exhaustive (how could it be?), but rather is intended to stimulate readers to envision for themselves new challenges and opportunities.  

In drafting these challenges, we were strongly influenced by DiVincenzo's discussion of dogma and heresy in quantum computing \cite{DiVincenzo:2001fk}; in particular the DiVincenzo's assertion that ``We should treat [the quantum] canon the way physicists usually treat a canon, by criticizing, reinterpreting and creatively flouting it.''  

Moreover, a strong engineering tradition is not to criticize, reinterpret, or creatively flout canons, so much as to \emph{repurpose} them; thus a main focus of our five challenges is to repurpose ideas whose foundations in mathematics and science are reasonably well-established, in service of practical engineering objectives.

\begin{figure}
\centering%
\vspace{-0.1250\baselineskip}
\includegraphics[width=0.98\columnwidth]{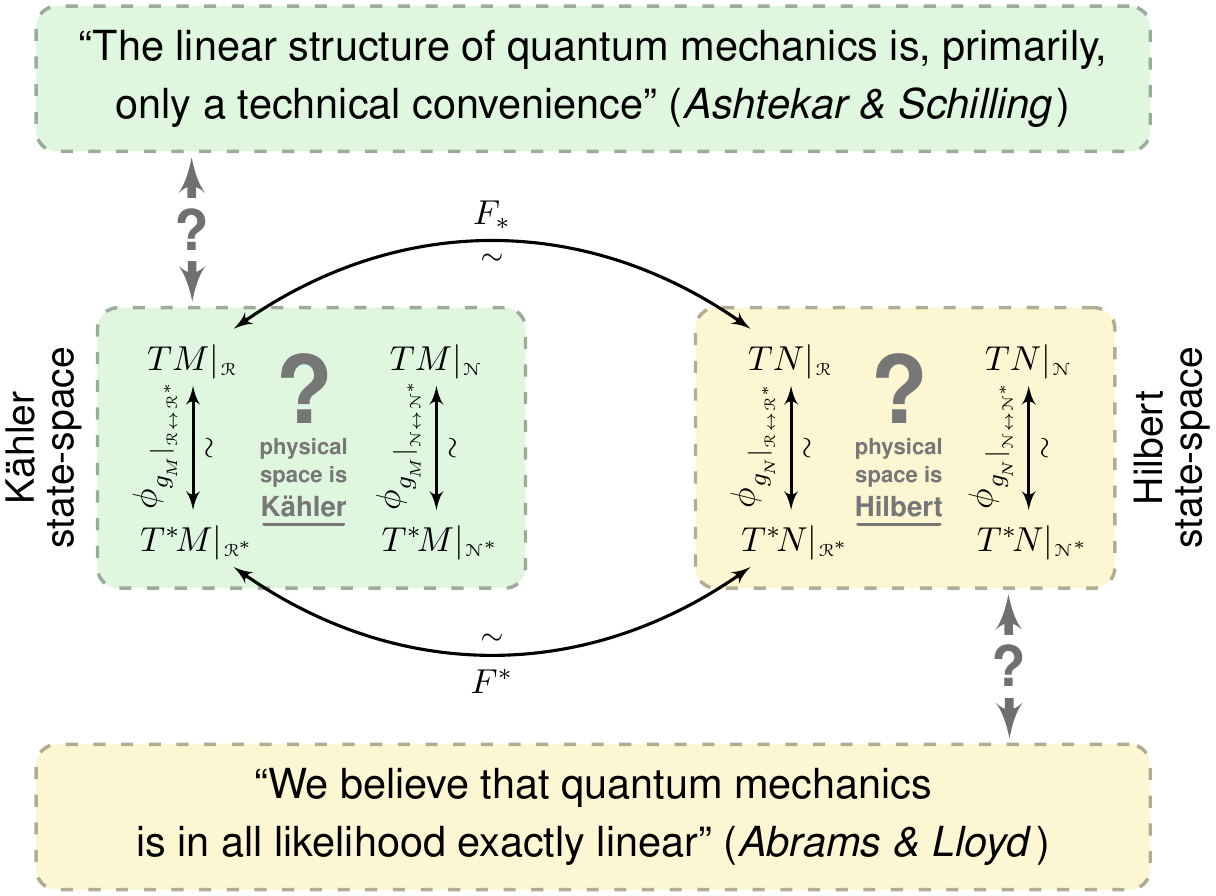}%
\vspace{-0.5\baselineskip}%
\caption[Hilbert space perspectives]{\label{fig: Feynman}%
Hilbert space from a simulationist perspective.  Is it feasible to follow Ash\-te\-kar and Schilling in regarding Hilbert space as (\onlinecite[][p.~61]{Ashtekar:1999yu}) ``primarily, only a~technical convenience''? %
}%
\end{figure}

\subsection{Build quantum-limited microscopes}

In two previous articles (one long \cite{Sidles:2009cl} and one short \cite{Sidles:2009yq}) we have reviewed the historical context and the scientific rationale of the following challenge: 

\myChallengeBox{\mbox{Quantum-limited spin microscopy}}%
{\label{challenge: microscopy}Provide first the mathematical and theoretical foundations, then the engineering design, then the committed enterprise, and finally the practical reality, of spin microscopes capable of observing all supramolecular structures, in three dimensions, with a voxel resolution of one nanometer or less.}

The new capability that natural simulation methods bring to this challenge is envisioned to be (with further development) the ability to reliably simulate the entire state-space of practical quantum spin microscopes.  It is hoped that this capability will contribute to an emerging enterprise that is known by many different names---synthetic biology, systems biology, and regenerative medicine, to name three---that is in reality the single large enterprise of achieving a working understanding of our own human biology, and our planet's.  Foreseeably this will be the largest single scientific enterprise of the 21st century; certainly it promises to be larger than any previous scientific enterprise of any previous century.

\subsection{Compute fermionic simulations efficiently}

The natural simulation framework developed in this article does not extend to the strongly-coupled fermionic systems that characterize much of materials science and quantum chemistry.  This is not an innate defect, but rather reflects mainly the authors' limited assimilation (to date) of the vast literature on this subject, as well as the practical difficulty of translating methods that often are empirical into the language of geometry and symplectic dynamics.  The following challenge is therefore suggested:

\myChallengeBox{\mbox{Efficient fermionic simulations}}%
{\label{challenge: isomorphism}Find efficient, general algorithms for calculating the musical isomorphism (that is, $T^{\ast\!}M{\leftrightarrow}TM\!$) on Grassmannian joins (in chemists' language, on joins of Slater determinants).}

Challenge \ref{challenge: microscopy} and \ref{challenge: isomorphism} are mutually reinforcing, in the sense that advances with respect to either substantially advance the other; this is especially evident in the rapidly emerging field of synthetic biology \cite{Jiang:2008ve,Rothlisberger:2008qy,Tantillo:1998bv}.

Challenge \ref{challenge: isomorphism}---and to a considerable extent all the challenges of this section---also is inspired largely by the seminal work of Ashtekar and Schilling \cite{Ashtekar:1999yu}, in particular their explicit recognition of the desirability of a spin-statistics postulate that ``refers only to the essential geometric structures'' and more broadly, their recognition of the need for ``viable, non-trivial generalizations of quantum kinematics for which even the measurement theory could be developed in detail.''

\subsection{Establish Lindbladian naturality}

A conspicuous gap on the overall naturality of our simulation framework is that the functional form of the Lindbladian potentials of Table~\ref{table: recipes} has at present no geometrically natural derivation.  Instead we derived these potentials by a tedious algebraic reduction from the Hilbert space expressions presented in Nielsen and Chuang~\cite{Nielsen:00}, guided by the physical ideas of our own recent \emph{Practical Recipes} article \cite{Sidles:2009cl}.  More naturally, these Lindbladian potentials would be constructed solely from intrinsic considerations of geometric, dynamical, and informatic naturality, that is, without reference to an embedding Hilbert space and thus without assuming \emph{a priori} that the state-space of quantum mechanics is a vector space.  

Despite efforts, we have to date not found such a natural construction for Lindbladian dynamical potentials.  We therefore pose this challenge:

\myChallengeBox{\mbox{Establish Lindbladian naturality}}%
{\label{challenge: Lindbladian}Construct a complete set of Lindbladian stochastic potentials solely from intrinsic considerations of geometric, dynamical, and informatic naturality (that is, without reference to an embedding Hilbert space) and prove that the associated communication channels are causal.}

Challenge~\ref{challenge: Lindbladian} is intended as a 21st century quantum analog to the 19th century challenge of evolving from a static Euclidean conception of geometry to a dynamical Riemannian conception.  The latter evolution can be broadly conceived as proceeding in four stages: Bowditch's 1807 discussion of the principles of differential geometry is startling modern (\onlinecite[][see p.~100]{Bowditch:1807fk}); Gauss's celebrated \emph{Theorema Egregium} of 1827 translated these ideas into intrinsic terms \cite{Gauss:1827}; by 1888 Riemann's multidimensional generalization of these ideas work was widely known and praised \cite{Baynes:1888fk}; and yet only in the 20th century was it experimentally demonstrated that that the geometry of space-time was Riemannian, relativistic, and dynamic.

\subsection{Demonstrate quantum oracles}%
\label{subsec: Demonstrate quantum oracles}

History teaches us that the evolution from Euclidean conceptions of geometry to Riemannian conceptions aroused passionate debate, and the same is already true of quantum geometry.  As illustrated in Fig.~\ref{fig: Feynman}, presently it is an open question whether tensor network state-spaces are best regarded as a dimensional reduction of Hilbert space induced by Lindbladian flow, or conversely, Hilbert space might perhaps be merely a dimensional augmentation of the \Kahlerian\ state-space of nature, imposed by 20th century physicists as a tool for simplifying calculations. 

To focus this debate productively, we therefore pose the following challenge:

\myChallengeBox{Demonstrate quantum oracles}
{\label{challenge: oracles}Experimentally demonstrate any oracle device that cannot be simulated in {\upshape\sffamily \scalebox{0.9}{PTIME}} (that is, with polynomial resources), or alternatively, provide experimental evidence that the geometric dynamics of~nature's quantum state-space obstructs such a demonstration.}

Here the idea is to broaden the scope and mitigate the technical challenges of quantum information science, by embracing some of the ideas of naturality that are associated to modern complexity theory. \footnote{We acknowledge the strong influence of recent work by Scott Aaronson and Alex Arkipov relating to quantum oracle capabilities \protect\cite{Aaronson:2010kx}, without suggesting thereby any implied endorsement by these authors of the key motivation of Challenge~\protect\ref{challenge: oracles}, namely, that quantum oracles serve as experimental probes of non-Hilbert quantum state-space geometry.}

To illustrate some of the considerations that arise in connection with Challenge~\ref{challenge: oracles}, let us consider recent work benchmarking control methods in 12-qubit spin systems  \cite{Negrevergne:2006uq}.  Direct simulation of a 12-qubit experiment requires a Hilbert space of dimension $\dim \lcal{H} = 2^{12}\,{=}\,4096$, which approaches the practical limit of simulation by density matrix methods \cite{Veshtort:2006fk,Bak:2000zp}.  On the other hand, we know from the work of Menicucci and Caves \cite{Menicucci:2002kx} that experiments in this class in principle can be explained with no quantum entanglement whatsoever.  The following question thus remains unanswered at present: what Schmidt rank is required for an accurate trajectory-based simulation of a 12-qubit system on a tensor network state-space?   The simulation framework described in this article is well-suited to answering this kind of question.

We can extend this question to ask, is there \emph{any} experiment we can accomplish, whose accurate simulation provably is infeasible with {\upshape\sffamily \scalebox{0.9}{PTIME}} computational resources?  

In our view, such an experiment would be comparable in scientific signifiance to Galileo's (possibly legendary) ball-drop from the Tower of Pisa.  Challenge~\protect\ref{challenge: oracles} is framed with a view toward helping to catalyze such an experiment.

It is tempting to frame these issues as a 21st century struggle between B{\oe}otian believers in a Hilbert state-space  \cite{Baynes:1888fk}, versus skeptics of quantum computing \cite{Aaronson:2004fk}.  A more traditional and sober view, though, is that a sustained tension between quantum B{\oe}otians and quantum skeptics is desirable as a stimulus to progress in quantum information science.  As ever, strong experiments analyzed in the light of well-posed theories have the final say.

\subsection{Create natural field theories}

The following question originated in discussions within our Quantum Systems Engineering (\QSE) Group:  Why is our universe friendly to simulation?

\myChallengeBox{\mbox{Naturality in field theory}}%
{\label{challenge: field theory}Construct a relativistic field theory whose near-field dynamics exhibits Lindbladian naturality, whose communication channels are provably causal, and whose quantum and space-time structures (that is, whose \Kahlerian\ and Riemannian structures) are dynamically coupled.}

As with the preceding challenges, Challenge~\ref{challenge: field theory} too has a dual aspect.   Even if it should happen that the quantum state-space of Nature is non-dynamical and perfectly Hilbert, there still may be substantial computational advantages attendant to regarding quantum state-space as dynamical and \Kahlerian; here the point is that efficient techniques for simulating quantum field dynamics has many practical uses in optics, radar, and condensed matter physics.

On the other hand, Challenge~\ref{challenge: field theory} encourages us to conceive of field theory along lines that differ substantially from 20th century conceptions, in which information-theoretic and complexity-theoretic considerations are similarly important as dynamical considerations.  We can envision, for example, an outcome along the lines ``$M$-theory is the unique, causally separable, relativistically invariant quantum field theory that can be simulated with {\upshape\sffamily \scalebox{0.9}{PTIME}} computational resources.''  

%
%
%
%
%
%


\section{Conclusions}
\label{sec: Conclusions}
%
%
%
%

The elements of naturality that we have described have allowed us to naturally specify a dynamical simulation framework that encompasses Hamiltonian, Lindbladian, and thermostatic flows on classical and quantum state-spaces.  

\subsection{Educational considerations}

Naturality is purchased at significant price: our description employs approximately one hundred abstract mathematical terms over-and-beyond those terms that typically are taught in the context of an undergraduate education in mathematics, engineering, and science.   Excellent textbooks exist for learning these elements of naturality---such the dozen canonical texts \myCiteCanonical\ suggested in Sec.~\ref{subsec: Dovetailed elements of naturality}---but at present there is no textbook that presents these naturality concepts in the unified context of practical simulation; perhaps someday textbooks on simulation may be written taking this perspective.


Once the canons of naturality are mastered, the simulation framework presented in this article itself flows naturally from elementary considerations of geometry, dynamics, and information theory.  Only a handful of equations need be grasped, and key proofs are associated to to projective diagrams whose form is easily recalled (see Theorems 1--3 of Sec.~\ref{sec: The summit: valid design criteria} and Tables~\ref{table: metric flow}--\ref{table: symplectic flow}).

\subsection{Unleashing capabilities from tethers}

The five natural simulation challenges whose technical aspects are discussed in the preceding section (Section \ref{sec: Five natural challenges}) can be understood as being collectively focussed on repurposing existing  capabilities and creating new capabilities that unleash us from present-day tethers, as follows.

\mysubsubsection{Challenge~\ref{challenge: microscopy}: Quantum-limited spin microscopy}
The spin microscopy challenge seeks to unleash us from the tether of pursuing structural and synthetic biology, and concomitantly regenerative biomedicine, primarily by experimental methods. As our recent overview of spin microscopy concludes:\cite{Sidles:2009yq} 
\begin{quote}We are tantalized by a vision of medical practice becoming fully curative and regenerative. We are frustrated---as the generation of von Neumann and Wiener was frustrated---by the limitations of our present tools. We desire---as Feynman famously desired \cite{Feynman:59}---to ``just look at the thing''. And we plan---as every previous generation has planned---for these aspirations to become realities.\end{quote}

The present article is thus the third and final article in a tryptych \cite{Sidles:2009yq,Sidles:2009cl} that focusses on von Neumann's challenge \cite{Sidles:2009cl} to consider whether atomic-resolution biological microscopy might be achieved ``by~developments of which we can already foresee the character, the caliber, and the duration. And are the latter two not excessive and impractical?''  

The natural simulation framework presented in this article was conceived, first and foremost,  to help 21st century mathematicians, scientists, and engineers to meet von Neumann's biomicroscopy challenge, and thereby to dynamically augment of knowledge of healing, by creating a balanced surge of new biomedical research capabilities.

\mysubsubsection{Challenge~\ref{challenge: isomorphism}: Efficient Fermionic simulation}

The Fermionic simulation challenge seeks to unleash condensed matter physics from the tether of Fermi liquid models, by providing efficient numerical tools for exploring condensed matter physics beyond Fermi-type approximations.  This challenge seeks to bring to materials science a predictive power comparable in scope to the observational power that quantum spin microscopy seeks to bring to biomedical science.  From a practical point of view, this augmentation of materials science capability will play a strategically vital role in the 21st century, as spurred by urgent necessity, humanity seeks new ways to free itself from age-old tethers like carbon-burning and resource shortfalls.

\mysubsubsection{Challenge~\ref{challenge: Lindbladian}: Establish Lindbladian naturality}

The Lindbladian naturality challenge seeks to expand the predictive power of quantum physics by unleashing it from the tether of Hilbert space.  No matter whether the state-space of Nature is a Hilbert space or not, by recognizing that Lindbladian flow is generically dimension-reducing, we gain immense new computational powers and design insights in the quantum realm.

\mysubsubsection{Challenge~\ref{challenge: oracles}: Demonstrate quantum oracles}

The quantum oracle challenge seeks to expand the realm of quantum information science by unleashing quantum information processing from the tether of quantum computing.   The key recognition is that the quantum oracle challenge is mathematically broader and potentially easier of experimental achievement than quantum computing, and yet is similarly seminal in its influence on our understanding of the computational aspects of our universe. 

\mysubsubsection{Challenge~\ref{challenge: field theory}: Naturality in field theory}
This mathematically toughest of our five challenges seeks to expand the capabilities of field theory by unleashing it from the tether of dynamical dominance.  The key recognition is that informatic flow is equally central to dynamical flow, and that the quantum and space-time structures of nature's state-space are (potentially) naturally dynamic and mutually coupled.

We close by emphasizing that the above five challenges (and many more that might be conceived) are closely dovetailed, in the sense that progress on any one of them acts to speed the pace and retire the risks of progress toward meeting all of them.

\vspace{\baselineskip}

\begin{acknowledgements}
\noindent This work was supported by the Army Research Office (ARO) MURI program under contract {\#}\,W911NF-05-1-0403. Prior support was provided by the ARO under contact {\#}\,DAAD19-02-1-0344, and by the NSF/ENG/ECS under grant {\#}\,0097544.\par
\vspace{0.65\baselineskip}
\noindent This work is dedicated to the families of the \emph{Ceremony in Honor of Wounded Marines}, 12 May 2006, Marine Corps Barracks, Washington, DC.
\end{acknowledgements}

\onecolumngrid

\newpage
\twocolumngrid

\appendix
\begin{table*}
\vspace*{-0.5\baselineskip}%
\caption[Recipes for Lindbladian dynamics]{\label{table: recipes}Recipes for the simulation of Lindbladian quantum dynamics on \Kahlerian\  state-spaces.}%
\centering%
\scalebox{1.0}[0.99]{\makebox[\textwidth][c]{%
\includegraphics[width=1.0125\textwidth]{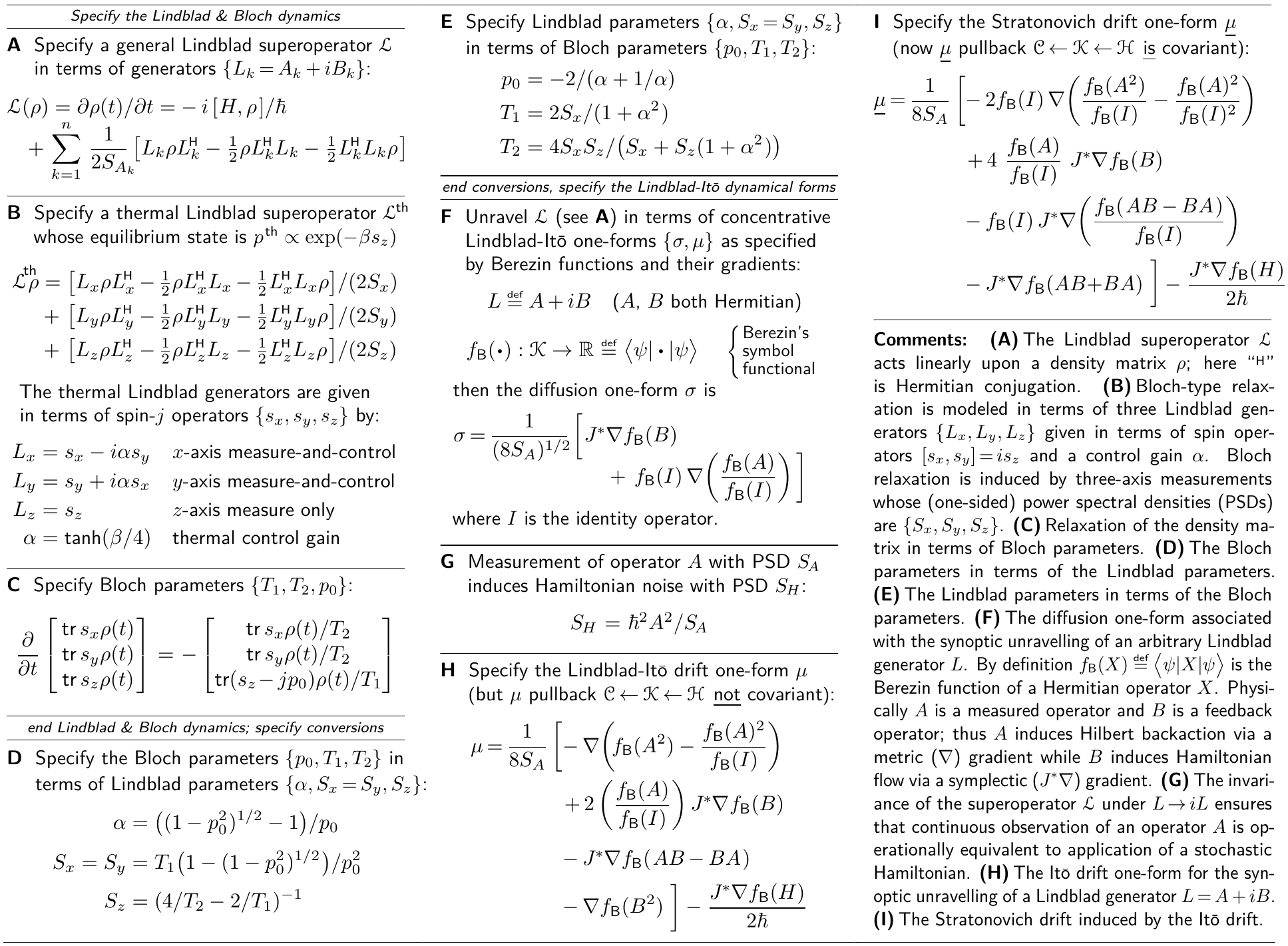}}}%
\vspace{-1.25\baselineskip}
\end{table*}

\section{Lindbladian recipes}
\label{sec: Lindbladian recipes} 
 
Given the general \emph{Bloch parameters} of a quantum spin system, this supplement provides explicit expressions for that system's \Kahlerian\ Lindblad-\Ito-Stratonovich pullback (\KLISP) one-forms \myTableThreeEntry{I}.  These expressions suffice to code the simulations of the main article, and more broadly, they suffice (in principle) to simulate all quantum systems whose dynamics are Lindbladian, including classical systems that respect the \emph{standard quantum limits} (\onlinecite[see Secs.~3.2.8 and 3.4 of][]{Sidles:2009cl}).

%
%

Three key elements of geometric simulation methods are summarized in this appendix: first, a Bloch-complete tabulation of Lindblad-\Ito\ generators \myTableFourEntry{A--E}, second, a general covariant (Stratonovich) form of the \KLISP\ increment \myTableFourEntry{G--I}, and third, a discussion of practical implications of \Kahlerian\ metric factoring. 

\subsection{Completely positive linear maps}
\label{subsec: Completely positive linear maps}
 
To begin, it is well known (\onlinecite[][see Sec.~8.4.1]{Nielsen:00}) that the most general infinitesimal generator of a completely positive linear map on a density matrix $\rho$ (physically, a map that preserves the positivity of probabilities) is of the following Lindbladian form:  
\begin{multline}
\label{eq: Lindbladian}
\lcal{L}(\rho) =  -i [\hspace{0.1em}H,\rho\hspace{0.1em}]/\hbar 
\\
 + \sum_{k = 1}^n \frac{1}{2S_{A_k}\hspace{-0.5em}}\left[\,\rule{0pt}{2.5ex}L_k\rho L^{\myscriptHermitian}_k
-\tfrac{1}{2} \rho L^{\myscriptHermitian}_k L_k
-\tfrac{1}{2} L^{\myscriptHermitian}_k L_k \rho\,\right],
\end{multline}
where $H$ is a (Hermitian) Hamiltonian operator, the $\{L_k;\ k\in1,\ldots n\}$ are a set of (nonHermitian) \emph{Lindbladian generators}, and a superscript ``\raisebox{0.35ex}{\mbox{$\,{\myscriptHermitian}\,$}}'' indicates Hermitian conjugation.   Without loss of generality we take  $L_k = A_k + i B_k$, with $A_k$ and $B_k$ being Hermitian operators and we introduce a (real, positive) scalar normalizing factor $S_{A_k\!}$ that we will identify with a  (one-sided) measurement noise spectral density having units of $A^2\cdot\text{time}$.  

Note that the normalizing factor $S_{A_k}$ in Eq.~\ref{eq: Lindbladian} allows the operators $L_k$ to be assigned any desired physical units (for example, position, linear momentum, angular momentum).

\subsection{Quantum operation unravellings}
\label{subsec: Quantum operation unravellings}

We focus our attention upon the Lindbladian map that is generated by a single (arbitrary) Lindbladian operator $L$. Following Nielsen and Chuang %
(\onlinecite[][chs{.}\,2 and 8]{Nielsen:00}) %
we first exhibit a \emph{stochastic unravelling} of a simulated wave function $\psi(t)$ that is specified in terms of \emph{quantum operations} $M^+$ and $M^-$ by
\begin{equation}
\label{eq: stochastic}
\psi(t+{\delta}t) = %
\left\{
\begin{array}{@{\hspace{0.1em}}l@{\quad}l}
M^{+}\psi(t)/\sqrt{p^+}&\text{with probability $p^+$,}\\[1ex]
M^{-}\psi(t)/\sqrt{p^-}&\text{with probability $p^-$,}
\end{array}
\right.
\end{equation}
where the probabilities $p^\pm$ are given by
\begin{equation}
\label{eq: probabilities}
p^\pm = \inner{M^{\pm}\psi(t)}{M^{\pm}\psi(t)},
\end{equation}
and the form of the measurement operations $M^{\pm}$ is
\begin{equation}
\label{eq: operations}
\begin{aligned}[b]
M^{\pm} = &\exp\big(-i(AB+BA){\delta}t/(4S_A)\big) 
\\
&\times \exp\big(\pm i B\sqrt{{\delta}t/(2S_A)\!}\,\,\big) %
\\
&\times \big[\cos\big(A\sqrt{{\delta}t/(2S_A)\!}\,\,\big)\\
&\qquad \pm
\sin\big(A\sqrt{{\delta}t/(2S_A)\!}\,\,\big)\big]/\sqrt{2}. 
\end{aligned}
\end{equation}
By the physical reasoning of \cite[Sec.~3.3 and Fig.~9]{Sidles:2009cl}, we recognize Eq.~(\ref{eq: operations}) as a synoptic unravelling, with $A$ as the measured operator, $B$ as the feedback Hamiltonian operator, and the term $\propto (AB+BA)$ as a Stark shift Hamiltonian.

The time-evolution of $\rho$ then is given in terms of the $M^\pm$ by 
\begin{equation}
\begin{aligned}[b]
\label{eq: M evolution}
\rho(t+{\delta}t)-\rho(t) =\ & M^{+}\rho(t) (M^{+})^{\myscriptHermitian} \\[1ex]
& + M^{-}\rho(t) (M^{-})^{\myscriptHermitian}-\rho(t).
\end{aligned}
\end{equation}
Substituting Eq.~(\ref{eq: stochastic}-\ref{eq: operations}) in Eq.~(\ref{eq: M evolution}) and expanding in $\smash{\sqrt{\delta t}}$ immediately establishes the equivalence of the quantum operation unravelling to the Lindbladian map:
\begin{align}
\lcal{L}(\rho) & = 
\lim_{\delta t\to 0} \frac{\rho(t+{\delta}t)-\rho(t)}{\delta t} \\
& = \frac{1}{2S_A}\,\big[
L\rho(t) L^{\myscriptHermitian}
-\tfrac{1}{2} \rho(t) L^{\myscriptHermitian} L
-\tfrac{1}{2} L^{\myscriptHermitian} L \rho(t) 
\big]%
.
\label{eq: single Lindbladian}
\end{align}

It is important to appreciate that the above Lindbladian map does not uniquely define the generator $L$; it is invariant under $L\,{\to}\, i L$, which corresponds to $A\,{\to}\,B$ and $B\,{\to}\,{-}A$. The operator $A$ is privileged in the unravelling by associating it with a measurement process; in consequence it is solely $A$ that will appear in the crucial (for purposes of concentration) variance-reducing drift one-forms of \myTableFourEntry{G} and \myTableFourEntry{I}.

\subsection{Finite-polarization Bloch equations}
\label{subsec: Finite-polarization Bloch equations}

Recipes for elaborating the above Lindbladian map into general-purpose recipes for simulating spin systems described by Bloch-type relaxation are well covered in the literature (\onlinecite[Secs.~3.4\,\&\,4.1][]{Sidles:2009cl}), and \myTableFourEntry{B--E} summarize the needed relations.  

In brief, Bloch-type relaxation to thermal equilibrium is unravelled as a process of measurement and control, in which the environment monitors the spin direction and applies control feedback necessary to establish the polarization is attendant to thermal equilibrium.   

More specifically, the operation representation of \myTableFourEntry{B} unravels  the thermalizing process as the sum of three measure\-ment-and-con\-trol operations, in which the expectation value of each of spin operators $\{s_x,s_y,s_z\}$ is continuously measured by processes having spectral densities $\{S_x,S_y,S_z\}$ (one-sided), and feedback control is applied with gain $\alpha$.  To assign physical values to the four thermalizing parameters $\{S_x,S_y,S_z,\alpha\}$, we first specify $S_x = S_y$ (transverse isotropy).  Then by a purely algebraic computation that depends only upon the spin commutation relations, we find that setting $\alpha = \tanh(\beta/4)$ ensures $\lcal{L}\big(\exp(-\beta s_z)\big) = 0$,  \emph{i.e.}, the  thermal density operator $\rho_\text{th} = \exp(-\beta s_z)$ is a fixed point of the thermalizing Lindbladian map.  

In this picture the control gain---the parameter $\alpha$ in \myTableFourEntry{B--E}---sets the temperature, with optimal gain of $\alpha\,{=}\,1$ corresponding to zero temperature, and lesser gains (or larger gains, see \cite{Sidles:2009cl}) corresponding to finite temperature.  

This physical picture explains why we cannot simulate quantum computations via geometric methods: the covert observation of the thermal environment destroys the high-order correlations upon which quantum computation depends, and which otherwise would be costly to simulate.  Present-day designs for quantum computing rely on an external supply of ancilla bits to error-correct these high-order correlations; these bits represent a steady increase in the dimension of the Hilbert space that similarly is not accommodated in the tensor network manifold upon which \KLISP increments are pulled-back.

The relations of \myTableFourEntry{B--E} are valid for arbitrary spin-$j$, and in particular the limit $j\,{\to}\,\infty$ describes the motion of test-masses; this limit too is discussed in-depth in \cite[Secs.~3.4\,\&\,4.1]{Sidles:2009cl}.

\subsection{Integrating stochastic equations}
\label{subsec: Integrating stochastic equations}
 
By a straightforward calculation, we can check that the synoptic unravelling of Eqs.~(\ref{eq: stochastic}--\ref{eq: M evolution}) is represented by the \Ito\ increment one-forms of \myTableFourEntry{F--H}.  However, in order to pullback these one-forms onto a lower-dimension $\lcal{K}$-manifold and then numerically integrate the trajectory, we must calculate a correction term called the \emph{Stratonovich correction}.  

We now derive this correction.   We introduce \Kahlerian\ indices as in the main article such that the Hilbert metric $g_\scriptH$ is given in terms of the \Kahler\ potential $\kappa(\psi)=\inner{\psi}{\psi}/2 = f_\text{B}(I)/2$ by
\begin{equation}
	[\hspace{0.05em}\raisebox{0.2ex}{$g_\scriptH$}\hspace{0.05em}]_{ab} 
	=
		G_\scriptH\!\left(\frac{\partial\hspace{0.25em}}{\partial{\psi}^{a}},\frac{\partial\hspace{0.25em}}{\partial\psi^{b}}\right) 
	= 
\frac{1}{2}\,\frac{\partial^2}{\partial\psi^a\partial\psi^b} f_\text{B}(I),
\end{equation}
where $f_\text{B}$ is \emph{Berezin's symbol functional} (see \myTableFourEntry{F} for definition) and $I$ is the identity operator on the Hilbert space $\lcal{H}$.  

A technical point is that our definition of Berezin's symbol functional is less restrictive than that of most authors in geometric quantum mechanics; for us $f_\text{B}$ is simply a linear map from Hermitian operators to real-valued state-space functions.  This reflects the fact that in our formalism Hamiltonian flow is a symplectomorphism (the Lie derivative of the symplectic form is zero because the pullback of a closed symplectic form is closed) but not necessarily a (metric) isomorphism (a flow that leaves the metric form invariant; compare Proposition 1.3 of \cite{Tyurin:2007dn}).   

We do not require that pulled-back Hamiltonian flow be a (metric) isomorphism because we know from prior theoretical work \cite[see Sec.~2 of][]{Sidles:2009cl} that tensor network state-spaces have nonuniform Riemannian curvature structure, such that Hamiltonian trajectories span regions of large and small state-space curvature.  Furthermore, as discussed in the main article, we know that symplectomorphic flow is a strong enough condition to ensure thermodynamically reasonable simulation physics.

As in the main article we define the raised tensor components $[\hspace{0.05em}\raisebox{0.2ex}{$g_\scriptH$}\hspace{0.05em}]^{ab}$ to be the (unique) matrix pseudo-inverse of $[\hspace{0.05em}\raisebox{0.2ex}{$g_\scriptH$}\hspace{0.05em}]_{ab}$.
\begin{equation}
	[\hspace{0.05em}\raisebox{0.2ex}{$g_\scriptH$}\hspace{0.05em}]^{ab}\,[\hspace{0.05em}\raisebox{0.2ex}{$g_\scriptH$}\hspace{0.05em}]_{bc} \,[\hspace{0.05em}\raisebox{0.2ex}{$g_\scriptH$}\hspace{0.05em}]^{cd}
	= 
	[\hspace{0.05em}\raisebox{0.2ex}{$g_\scriptH$}\hspace{0.05em}]^{ad}.
\end{equation}
The metric $g_\scriptK$ on the pullback \Kahler\ state-space $\lcal{K}$ is
 \begin{equation}
 \label{eq: metric}
 [\hspace{0.05em}\raisebox{0.2ex}{$g_\scriptK$}\hspace{0.05em}]_{ab} = \frac{\partial \psi^c}{\partial \xi^a} [\hspace{0.05em}\raisebox{0.2ex}{$g_\scriptH$}\hspace{0.05em}]_{cd} \frac{\partial \psi^d}{\partial \xi^b}. 
\end{equation}
Now we can define a (local) projection matrix $[\hspace{0.1em}P_\scriptK]^{a}_{\ b}$ in terms of the metric
on our state-space:
\begin{equation}
 [\hspace{0.1em}P_\scriptK]^{a}_{\ b} = %
 \frac{\partial\psi^a}{\partial\xi^m}[\hspace{0.05em}\raisebox{0.2ex}{$g_\scriptK$}\hspace{0.05em}]^{mn}\frac{\partial\psi^c}{\partial\xi^n}
 [\hspace{0.05em}\raisebox{0.2ex}{$g_\scriptH$}\hspace{0.05em}]_{cb}.
\end{equation}
We consider the Hilbert-space \Ito\ increment from \myTableFourEntry{A} of the main article: 
\begin{equation}
\mydIto{\psi}^a = [\hspace{0.05em}\raisebox{0.2ex}{$g_\scriptH$}\hspace{0.05em}]^{ab} 
	\left[\,%
	{\mu}\!\left(\frac{\partial\hspace{0.25em}}{\partial{\psi^a}}\right)
		\mydIto{t}
	+\
	{\sigma}\!\left(\frac{\partial\hspace{0.25em}}{\partial{\psi^a}}\right)
		\mydIto{W}
	\,\right].
	\label{eq: starting Ito}
\end{equation}
Applying $P_\scriptK$ to Eq.~(\ref{eq: starting Ito}) and using the chain rule,  
we obtain a covariant expression for the \Ito\ increment projected onto the tangent state-space
\begin{align}
	[\hspace{0.1em}P_\scriptK]^{a}_{\ b}\,\mydIto{\psi}^b 
	= 
	 \frac{\partial\psi^a}{\partial\xi^m}[\hspace{0.05em}\raisebox{0.2ex}{$g_\scriptK$}\hspace{0.05em}]^{mn}
	\bigg[\,&%
		{\mu}\!\left(\frac{\partial\hspace{0.25em}}{\partial{\xi^n}}\right)
			\mydIto{t}\nonumber\\
		&\ +\
		{\sigma}\!\left(\frac{\partial\hspace{0.25em}}{\partial{\xi^n}}\right)
			\mydIto{W}
		\,\bigg].
		\label{eq: first increment}
\end{align}
Now we let the $\xi$ functions be processes with \Ito\ increments 
$\{\mu^a, \sigma^a\}$ that we wish to solve for
\begin{equation}
\mydIto{\xi}^\alpha = \mu^\alpha \mydIto{t}  + \sigma^\alpha \mydIto{W}.
\end{equation}
\Ito's Lemma then gives the $\psi$-increments in terms of $\{\mu^a, \sigma^a\}$ as
\begin{equation}
\label{eq: Lemma}
\mydIto{\psi}^b =  
\frac{\partial\psi^b}{\partial\xi^p}\,
\left[\hspace{0.1em}
\mu^p \mydIto{t}  + \sigma^p \mydIto{W}\hspace{0.1em}
\right]
+
\frac{1}{2}\,\frac{\partial^2\psi^b}{\partial\xi^r \partial \xi^s}\,
\sigma^r\sigma^s\mydIto{t}.
\end{equation}
Applying $P_\scriptK$ yields
\begin{align}
	[\hspace{0.1em}P_\scriptK]^{a}_{\ b}&\,\mydIto{\psi}^b  
	=\  
	\frac{\partial\psi^a}{\partial\xi^m}[\hspace{0.05em}\raisebox{0.2ex}{$g_\scriptK$}\hspace{0.05em}]^{mn}
	\bigg[[\hspace{0.05em}\raisebox{0.2ex}{$g_\scriptK$}\hspace{0.05em}]_{np}\Big[  
	\mu^p \mydIto{t}  + \sigma^p \mydIto{W}\nonumber\\
	&\ + \frac{1}{2}\,[\hspace{0.05em}\raisebox{0.2ex}{$g_\scriptK$}\hspace{0.05em}]^{pq}\frac{\partial\psi^c}{\partial\xi^n}[\hspace{0.05em}\raisebox{0.2ex}{$g_\scriptH$}\hspace{0.05em}]_{cb}\frac{\partial^2\psi^b}{\partial\xi^r \partial \xi^s}\,
	\sigma^r\sigma^s\mydIto{t}
	\Big]\bigg].
	\label{eq: second increment}
\end{align}
We equate the right-hand-sides of Eqs.~(\ref{eq: first increment}) and (\ref{eq: second increment}), and switch to \Kahlerian\ indices.  Recalling Eq.~(\ref{eq: metric}), we readily verify that the resulting equation is satisfied by the following $\xi^{\hspace{-0.1em}\alpha}$-coordinate \Ito\ drift and diffusion: 
\begin{align}
\sigma^\alpha & = [\hspace{0.05em}\raisebox{0.2ex}{$g_\scriptK$}\hspace{0.05em}]^{\alpha \sbar{\beta}}\,{\sigma}\left(\frac{\partial}{\partial\xi^{\sbar{\beta}}}\right),
\\
\mu^\alpha & = [\hspace{0.05em}\raisebox{0.2ex}{$g_\scriptK$}\hspace{0.05em}]^{\alpha \sbar{\beta}}\left[
{\mu}\left(\frac{\partial}{\partial\xi^{\sbar{\beta}}}\right)
-\frac{1}{2}\frac{\partial [\hspace{0.05em}\raisebox{0.2ex}{$g_\scriptK$}\hspace{0.05em}]_{\sbar{\beta}\gamma\rule[0pt]{0pt}{0pt}}}{\partial \xi^\delta}\,\sigma^\gamma \sigma^\delta\right].
\label{eq: stratonovich correction}
\end{align}
This expression has the computationally desirable property of being wholly intrinsic to the $\lcal{K}$-manifold.   However we notice that the pullback $\lcal{C}\,{\leftarrow}\,\lcal{K}\,{\leftarrow}\,\lcal{H}$ has induced a noncovariant correction to the \Ito\ drift (because the factor $\partial [\hspace{0.05em}\raisebox{0.2ex}{$g_\scriptK$}\hspace{0.05em}]_{\raisebox{0ex}[0pt][0pt]{\text{$\scriptstyle\sbar{\beta}\gamma$}}}/\partial \xi^\delta$ in \mbox{Eq.~(\ref{eq: stratonovich correction})} does not transform as a tensor).  

We see that \Ito's Lemma generically induces noncovariant drifts whenever \Ito\ increments are pulled-back onto nonEuclidean manifolds.   This implies that \Ito\ drift one-forms are not natural geometric objects; in the following section this will motivate us to define Stratonovich drift one-forms that \emph{are} geometrically natural.

However, \Ito\ increments are quite convenient for numerical computations, and the dynamic nuclear polarization (\DNP) trajectories reported in this article were in fact computed by integrating \Ito\ increments.  Remark: the high-precision ($\pm0.02$ in polarization) agreement between density matrix and geometric trajectory calculations that is seen in \myTableThreeEntry{C} required the inclusion of the correction term in Eq.~(\ref{eq: stratonovich correction}); this finding was contrary to our early expectations (or hopes) that the correction term might be dynamically negligible.

More broadly, the expressions collected in Table~3 of the main article, together with the compendium of Table~4, plus the above drift correction of \mbox{Eq.~(\ref{eq: stratonovich correction})}, suffice to numerically simulate any spin system whose symplectic dynamics are Hamiltonian, whose metric dynamics are Lindbladian, and whose relaxation parameters are Bloch-type.

\subsection{\Ito-to-Stratonovich conversion}
\label{subsec: Ito-to-Stratonovich conversion}

A well known, formally equivalent, and geometrically covariant alternative to \Ito\ integration is Stratonovich integration, whose \Kahlerian\ drift and diffusion we now derive. 

In essence Stratonovich methods embrace an alternative stochastic integration scheme that eliminates the noncovariant correction term in Eqs.~(\ref{eq: Lemma}) and (\ref{eq: stratonovich correction}), such that the drift and diffusion one-forms pullback $\lcal{C}\,{\leftarrow}\,\lcal{K}\,{\leftarrow}\,\lcal{H}$ covariantly, and thus retain a natural geometric interpretation. This geometric naturalness is obtained at the expense of a more complicated, nonMarkovian definition of stochastic integration; this trade-off is the subject of a large literature.  In general \Ito\ and Stratonovich integration methods {both} are widely used, and the decision of which to use is made on a case-by-case basis.  


\newcommand{\mymubar}{%
\ensuremath{\myunderset{\raisebox{1.3ex}[0pt][0pt]{\rule{1.75ex}{0.225ex}}}{\mu}}}

\renewcommand{\mymubar}{%
\ensuremath{\myunderset{\raisebox{1.75ex}[0pt][0pt]{\hspace{0.55ex}\rule{1.20ex}{0.20ex}}}{\mu}}}

\newcommand{\myunderset}[2]{%
	\raisebox{0ex}[0pt][0pt]{\mbox{$\displaystyle%
	\underset{%
	\raisebox{0ex}[0pt][0pt]{\mbox{%
	\text{\scalebox{1.40}{\mbox{\tiny\sffamily{#1}}}}}}}{#2}$}}}

On the Hilbert manifold $\lcal{H}$ the Stratonovich drift one-form $\mymubar$ is given covariantly in terms of the \Ito\ drift one-form $\mu$ by a well known conversion relation:
\begin{align}\nonumber
	\mymubar%
   \left(%
   \hspace{-0.1em}
   \frac{\partial\ }{\partial\psi^a} 
   \!\right) =\ & 
	\mu
   \left(\hspace{-0.1em}\frac{\partial\ }{\partial\psi^a}\!\right) 
   \\
   &\ -\frac{1}{2}\,
   [\hspace{0.05em}\raisebox{0.2ex}{$g_\scriptH$}\hspace{0.05em}]^{bc}\,\sigma\hspace{-0.2em}
   \left(\hspace{-0.1em}
   \frac{\partial\ }{\partial\psi^c} 
   \!\right)
   \frac{\partial\ }{\partial\psi^b}\left(
   \sigma\hspace{-0.2em}
   \left(\hspace{-0.1em}
   \frac{\partial\ }{\partial\psi^a} 
   \!\right)\right).
\end{align}
Switching to \Kahlerian\ indices, but still working on $\lcal{H}$, we obtain
\begin{alignat}{1}\nonumber
   \mymubar\left(%
   \hspace{-0.1em}
   \frac{\partial\ }{\partial\psi^{\sbar\alpha}}
   \!\right) =\ &
	\mu
   \left(\hspace{-0.1em}\frac{\partial\ }{\partial\psi^{\sbar\alpha}}\!\right) \\
   \nonumber
   \hspace{-2.5em}-\frac{1}{2}\,\Bigg[\,%
   &[\hspace{0.05em}\raisebox{0.2ex}{$g_\scriptH$}\hspace{0.05em}]^{\beta{\sbar\gamma}}\,\sigma\hspace{-0.2em}
   \left(\hspace{-0.1em}
   \frac{\partial\ }{\partial\psi^{\sbar\gamma}} 
   \!\right)
   \frac{\partial\ }{\partial\psi^{\beta}}\left(
   \sigma\hspace{-0.2em}
   \left(\hspace{-0.1em}
   \frac{\partial\ }{\partial\psi^{\sbar\alpha}} 
    \!\right)\right)\\
   +\ &[\hspace{0.05em}\raisebox{0.2ex}{$g_\scriptH$}\hspace{0.05em}]^{\sbar\beta{\gamma}}\,\sigma\hspace{-0.2em}
   \left(\hspace{-0.1em}
   \frac{\partial\ }{\partial\psi^{\gamma}} 
   \!\right)
   \frac{\partial\ }{\partial\psi^{\sbar\beta}}\left(
   \sigma\hspace{-0.2em}
   \left(\hspace{-0.1em}
   \frac{\partial\ }{\partial\psi^{\sbar\alpha}} 
    \!\right)\right) \Bigg].
\end{alignat}
Substituting the \Ito\ drift and diffusion from \myTableFourEntry{F} and \myTableFourEntry{H}, carrying through the indicated derivatives, and expressing the result in terms of Berezin functions and gradient one-forms, yields the Stratonovich drift one-forms of \myTableFourEntry{I}.   

We find that the Stratonovich drift of F \myTableFourEntry{I} is variance-reducing and thus concentrative, just like the \Ito\ drift of \myTableFourEntry{H}.

The chief practical advantage of Stratonovich drifts for simulation purposes is that they pullback covariantly onto the reduced-dimension state-space $\lcal{K}$; in particular the noncovariant (and computationally costly) correction terms of the preceding section are not induced by the pullback.  

\subsection{\Kahlerian\ metric factoring}
\label{subsec: Kahlerian metric factoring}

In our own research we have not as yet numerically experimented with Stratonovich integration; however Kloeden and Platen's textbook \cite[see Secs.~11.1 and 11.5 of][]{Kloeden:1992kx} suggests specific lines of inquiry for stochastic systems in this class.  For example, the state-space flows induced by the diffusive one-forms of \myTableFourEntry{F} commute in the stochastic sense if and only if their respective Lindbladian generators commute.   Numerical stochastic inte\-gration---both \Ito\ and Stratonovich---can be made more efficient whenever all (or most) diffusions commute; this diffusion structure occurs naturally in Bloch-parameter spin models. 

As was discussed in the main article, computing the gradient-to-vector isomorphism for a system of $n\,{=}\,500$ spin-{$1\hspace{-0.1em}/2$} particles, pulled-back onto a rank $r\,{=}\,100$ tensor network state-space, requires the solution of ${2nr}\,{=}\,10^5$ simultaneous equations.  Such systems are too large for direct numerical solution, and it is well known \cite{Greenbaum:1997nr,Loan:96,Barrett:1994nr} that iterative methods can lead to $\lcal{O}(n)$-fold speed-up whenever the associated matrix-vector product can be evaluated in $\lcal{O}(n)$ operations.  For systems having $n$ of order $10^2$ or larger, this can be the difference between feasible and infeasible simulation; for this reason iterative methods are almost universally employed in large-scale classical simulations.

Iterative methods, generally speaking, are optimized on a case-by-case basis; thus it is reasonable to anticipate that the development of iterative methods for integrating simulation trajectories will proceed by a blend of fundamental theory guided by practical computational experience.

\subsection{Observations upon trajectory concentration}
\label{subsec: Observations upon trajectory concentration}

Let us examine the expressions for \KLISP diffusion \myTableFourEntry{F} and Stratonovich drift \myTableFourEntry{I} with a view toward identifying aspects of trajectory concentration that are understood versus \emph{not} understood, in both mathematical and physical terms.

Turning our attention first to the diffusion one-forms \myTableFourEntry{F}, it is apparent that the symplectic gradient one-form $\propto \myJstar\nabla f_\text{B}(B)$ generates the Hamiltonian flow of the feedback noise, while the metric gradient one-form $\propto \nabla f_\text{B}(A)$ generates the \emph{Hilbert backaction} that is associated with the measurement of $A$.  It is this feedback that is dynamically responsible for familiar quantum phenomena such as the Stern-Gerlach effect, for example (see \cite[Sec.~3.2]{Sidles:2009cl}).

Turning our attention now to the drift one-forms \myTableFourEntry{I}, the sole metric gradient one-form is manifestly concentrative; it is responsible for the existence of the $P$-representations described in Sec.~3.2 of \cite{Sidles:2009cl}, for example.  The~symplectic gradient  $\propto \myJstar\nabla f_\text{B}(B)$ supplies the coherent feedback associated with measurement of $A$.  The commutator term $AB+BA$ (which functions as a Stark shift Hamiltonian) was inserted at the beginning (Eq.~\ref{eq: operations}) solely to obtain the conventional Lindbladian map (Eq.~\ref{eq: Lindbladian}).  The anticommutator term $AB-BA$ presumably has geometric significance, but this significance is opaque (to us at least).

The unravelling we have been discussing is synoptic, and thus the preceding discussion is informed by a physical picture in which $A$ is measured and $B$ specifies feedback.  However, the synoptic representation is not unique; there exists a unitary invariance---called the \emph{Theorema Dilectum} (theorem of choice)\cite{Sidles:2009cl}---which guarantees that a class of informatically equivalent symplectic and metric one-forms can be unravelled from any given starting Lindbladian map.   This theorem is manifest in the invariance of the Lindbladian generator $\lcal{L}$ of Eq.~(\ref{eq: Lindbladian}) under the mapping induced by $L_k \mapsto \sum_j U_{kj}L_j$, where $U$ is an arbitrary unitary matrix.  The~physical and informatic significance of the $U$-matrix is discussed in %
\cite{Sidles:2009cl}; in essence $U$-equivalence is necessary to the causal separability of quantum measurements. 
	
	Until we can survey the entire $U$-equivalence class of \KLISP one-forms with mathematical equanimity, it cannot be said that we have a satisfactory understanding of the geometric aspects of quantum dynamics.    Similarly, until we can select---or better, design---particular unravellings that are well-suited to efficient simulation,  it cannot be said that we possess even a working grasp of quantum simulation's computational complexity.%
	
	This uncertainty extends to simulation's close cousin, optimization.
	Suppose for example that we have a classical annealing algorithm for
	solving (inefficiently) an \NP-complete problem like 3-\SAT.  If we
	view the 3-\SAT bits as qubits, and the annealing process as a quantum
	annealing process, then we can ask questions like: ``Bearing in mind
	the gauge-like invariance of the \emph{Theorema Dilectum} that is associated to
	Lindbladian dynamics \cite{Sidles:2009cl}, what
	quantum noise model yields the most effective annealing on a Schmidt
	rank-$r$ state-space?''  Section~4.6 of \cite{Sidles:2009cl} discusses
	these questions, but informatically speaking, we know very little about
	optimizing flow on \Kahlerian\ state-spaces.

\bibliography{QSE_JCP}

\end{document}